\documentclass[iop]{emulateapj}
\usepackage{apjfonts}

\usepackage{hyperref}
\hypersetup{
    colorlinks=true,
    linkcolor=blue,
    filecolor=magenta,      
    urlcolor=blue,
    citecolor=blue
}
\slugcomment{Not to appear in Nonlearned J., 45.}

\shorttitle{Arcade Implosion Caused by a Filament Eruption in a Flare}
\shortauthors{Wang et al.}

\begin{document}

\title{Arcade Implosion Caused by a Filament Eruption in a Flare}

\author{Juntao Wang, P. J. A. Sim\~{o}es, and L. Fletcher}
\affil{SUPA, School of Physics and Astronomy, University of Glasgow, Glasgow G12 8QQ, UK}
\email{j.wang.4@research.gla.ac.uk}

\author{J. K. Thalmann}
\affil{Institute of Physics/IGAM, University of Graz, Universit\"atsplatz 5, A-8010 Graz, Austria}

\author{H. S. Hudson}
\affil{SSL/UC, Berkeley, CA, USA}

\and 

\author{I. G. Hannah}
\affil{SUPA, School of Physics and Astronomy, University of Glasgow, Glasgow G12 8QQ, UK}

\begin{abstract}
Coronal implosions - the convergence motion of plasmas and entrained magnetic field in the corona due to a reduction in magnetic pressure - can help to locate and track sites of magnetic energy release or redistribution during solar flares and eruptions. We report here on the analysis of a well-observed implosion in the form of an arcade contraction associated with a filament eruption, during the C3.5 flare SOL2013-06-19T07:29. A sequence of events including magnetic flux-rope instability and distortion, followed by filament eruption and arcade implosion, lead us to conclude that the implosion arises from the transfer of magnetic energy from beneath the arcade as part of the global magnetic instability, rather than due to local magnetic energy dissipation in the flare. The observed net contraction of the imploding loops, which is found also in nonlinear force-free field extrapolations, reflects a permanent reduction of magnetic energy underneath the arcade. This event shows t
hat, in addition to resulting in expansion or eruption of overlying field, flux-rope instability can also simultaneously implode unopened field due to magnetic energy transfer. It demonstrates the ``partial opening of the field'' scenario, which is one of the ways in 3D to produce a magnetic eruption without violating the Aly-Sturrock hypothesis. In the framework of this observation we also propose a unification of three main concepts for active region magnetic evolution, namely the metastable eruption model, the implosion conjecture, and the standard ``CSHKP'' flare model.
\end{abstract}

\keywords{Sun: atmosphere --- Sun: filaments, prominences --- Sun: flares --- Sun: magnetic fields --- Sun: UV radiation --- Sun: X-rays, gamma rays}

\section{Introduction}

A solar flare is a sudden brightening in the solar atmosphere \citep[for an observational review, see][]{fle2011}. Flares are believed to be caused by the release
of free magnetic energy, which can be represented as magnetic field shear or twist in the corona. The energy release, possibly triggered by an instability or nonequilibrium, heats or accelerates particles, and often leads to a large scale ejection. The means to track the onset of the instability , the movement of free energy through the corona, and the location of the energy release or conversion would significantly assist with efforts to understand and predict the conditions leading to a flare or eruption. The implosion conjecture, first proposed by \citet{hud2000}, may help. As magnetic energy release corresponds to magnetic pressure reduction, peripheral loops \citep{she2014} would experience convergence towards the energy release site in order to obtain a new equilibrium position. Overlying loops would contract (well illustrated in Figure 3 of \citeauthor{rus2015} \citeyear{rus2015}), whereas underlying loops, if they exist, may show expansion up towards the energy releas
e site \citep{she2012}. In this paper, we study the evolution of an active region (AR) in which a clear implosion in one part of the region takes place, along with a filament eruption and two flares. Supported by extrapolations of the coronal magnetic field, we interpret the observed sequence of events as due to the transfer of magnetic energy from one part of the AR to another, from which the eruption and energy release can take place.

There are only a few implosion events observed in the periphery of the AR. From 2009 to 2012, Liu and
other collaborators report a series of events showing coronal loop contractions in the extreme ultraviolet (EUV) \citep{liu2009b,liuw2009,liu2010,liu2012}. These events ranges from
GOES class B to X with contraction speeds from tens to hundreds of km/s, happening in the preflare phase, during the impulsive phase or in the gradual phase.
It seems that implosion is possible in all flare classes and during the entire flare process. Some authors observe loop contractions accompanying erupting filaments or bubbles \citep{liuw2009,liu2012,sim2013a,yan2013,she2014,kus2015}. \citet{sim2013a} in an M6.4 flare find that the loop contraction speed correlates well with the hard X-ray (HXR) and microwave (MW) radiation, with faster contraction corresponding to more intense radiation. For the X2.2
flare SOL2011-02-15T01:50, \citet{gos2012} reports a stepwise permanent decrease in the longitudinal photospheric magnetic field after the flare, with the
field becoming more inclined towards the polarity inversion
line (PIL);
\citet{sun2012} exploit vector magnetograms
and find a mean horizontal photospheric field enhancement of 28\%
compared to the preflare state. They argue that the more horizontal field after the flare is caused by the observed implosion of loops. Loop oscillation accompanying or following contraction has also been observed \citep{liu2010,gos2012,liu2012,sun2012,sim2013a}. \citet{rus2015} analyse these two kinds of motion of loops in theory and connect them as a single response to the underlying magnetic energy release. \citet{hud2000} also points out that the implosion process should be most pronounced in the impulsive phase when the energy release rate reaches its maximum. It should also be noted that the implosion conjecture is based on three assumptions \citep{hud2000}. They are as follows: (1) a flare or coronal mass ejection (CME) gets its energy directly from the solar corona;
 (2) gravitation is of no significance; and (3) there is a plasma $\beta\ll1$ in the corona. Thus regions not meeting the above assumptions may not exhibit implosion behaviours.

It is useful to distinguish two kinds of implosion. Many of the observations described above show convergence of loops at the edge of an AR towards its centre. We call this a ``peripheral implosion'' \citep{she2014} where the convergence of loops on the periphery of the region is the consequence of a central magnetic energy liberation but the change in the free energy of those peripheral loops is small compared to the flare energy. Such peripheral implosions are still
rarely reported, as introduced above. However, the energy-carrying core field could also implode. A ``core implosion'', involving the main energy storage region, would lead directly to the flare signatures of particle acceleration and radiation. For instance, the shrinkage of newly reconnected field lines caused by the enhanced magnetic tension force in the gradual phase \citep{for1996} could provide energy to flaring loops or a looptop source via, e.g., a collapsing magnetic trap  \citet{ver2006}, shocks \citep{lon2009}, or Alfv\'{e}n waves \citep{fle2008}. Thus if we want to generalise the implosion concept, reconnection may be taken as a special circumstance that could result in core implosion. \citet{ji2007} propose that the unshearing motion of magnetic field lines could also cause the field itself to implode.

The implosion conjecture links flare energy release with field contraction, and is apparently at odds with many flares in which eruptions are seen. The Aly-Sturrock hypothesis \citep{aly1984,aly1991,stu1991}, which states that the energy of any simply-connected and closed force-free field is less than the energy of the corresponding completely opened field with the same vertical flux at its boundary, implies that energy must be added to erupt the field, rather than being liberated by the process, as is required to explain the flare. One solution is the partial opening of the field in a three-dimensional (3D) configuration. Magnetohydrodynamic (MHD) simulations utilising the 3D metastable eruption model \citep{stu2001}, which has a twisted flux rope anchored below a magnetic arcade, have shown that during the flux rope eruption, some unopened overlying arcade loops could be pushed upward and aside and finally contract compared to their initial states \citep{rou2003,aul2005,gib
2006,fan2007,rac2009}. These are examples of peripheral implosion accompanying the central energy release manifested by the flux rope eruption. In our observations, implosion in one part of the AR accompanies an asymmetric eruption in another, reflecting energy transfer from the region of the imploding field to the erupting field.

This paper reports on a well-observed overlying arcade contraction associated with a filament eruption, which happen during two consecutive flares. We find evidence for implosion from observations and magnetic field extrapolations, and demonstrate the close
relationship between the two flares, the filament eruption, and the overlying arcade contraction. In Section~\ref{observation}, the observations of the entire event are described. In Section~\ref{extrapolation}, magnetic field extrapolations are exploited to reveal the implosion and possible reconnection between the filament and other AR field in the form of an extended ``arm-like'' structure. Discussion including possible scenarios for the evolution is presented in Section~\ref{discussion} and conclusions in \ref{conclusions}. 

\section{Observations} \label{observation}

\subsection{Overview of the Event}

\begin{figure}[]
\includegraphics[scale=0.69]{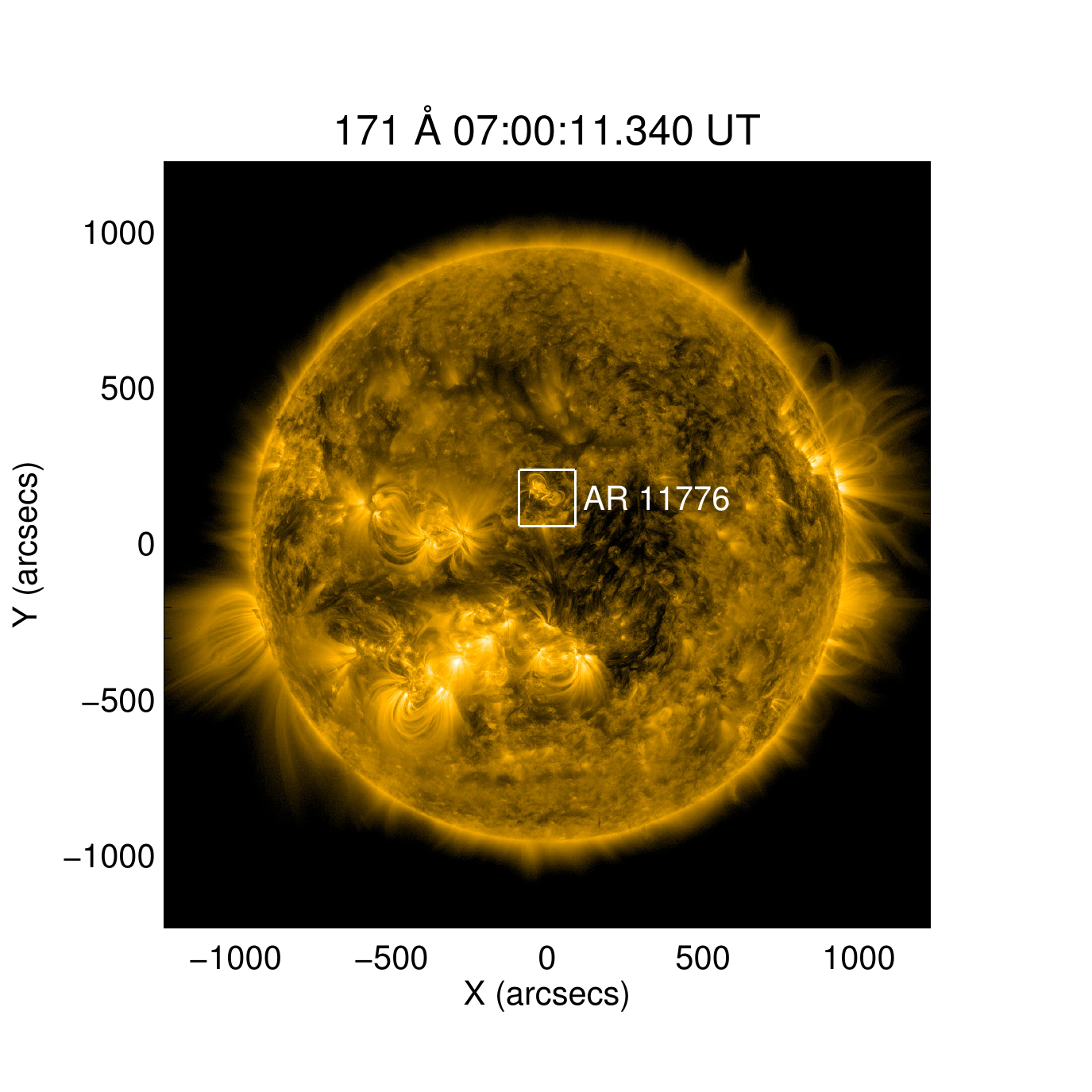}
\caption{\label{fullview} Full Sun image shows the AR 11776 on 2013 June 19 in AIA 171 {\AA}. The white square region is the field of view (FOV) used in Figure~\ref{structure}. A color version of this figure is available in the online journal.}
\end{figure}

On 2013 June 19, AR 11776 (N10W00) was located near the solar disk center (Figure~\ref{fullview}).  We focus our analysis around the period of the flare SOL20130619T07:29, GOES class C3.5. It was observed by the Reuven Ramaty High Energy Solar Spectroscopic Imager (RHESSI, \citeauthor{lin2002} \citeyear{liu2012}) and by the Solar Dynamics Observatory (SDO, \citeauthor{pes2012} \citeyear{pes2012}) instruments: Atmospheric Imaging Assembly (AIA, \citeauthor{lem2012} \citeyear{lem2012}) and Helioseismic and Magnetic Imager (HMI, \citeauthor{sche2012} \citeyear{sche2012}; \citeauthor{scho2012} \citeyear{scho2012}; \citeauthor{hoe2014} \citeyear{hoe2014}). The AIA images have been processed using standard software \citep{boe2012}, and also rotated to 07:00 UT via the drot\_map.pro procedure, in order to compensate for the solar differential rotation. RHESSI images were reconstructed using the CLEAN algorithm \citep{hur2002}, with detectors 2 to 8 and the \verb!clean_beam_width! set to 1.5 \citep{sch2007,sim2013b}. No CMEs associated with the event were reported, while a type III radio burst was detected\footnote{\url{http://secchirh.obspm.fr/survey.php?hour=0600&dayofyear=20130619&survey_type=4}} (for a review of type III radio bursts, see \citeauthor{rei2014} \citeyear{rei2014}), as well as an EUV wave, which will be briefly presented in Section~\ref{exstruc}. We note that earlier on the same day a C2.3 flare was produced in this AR and studied by \citet{zhe2015}.

\begin{figure}[]
\includegraphics[scale=0.95]{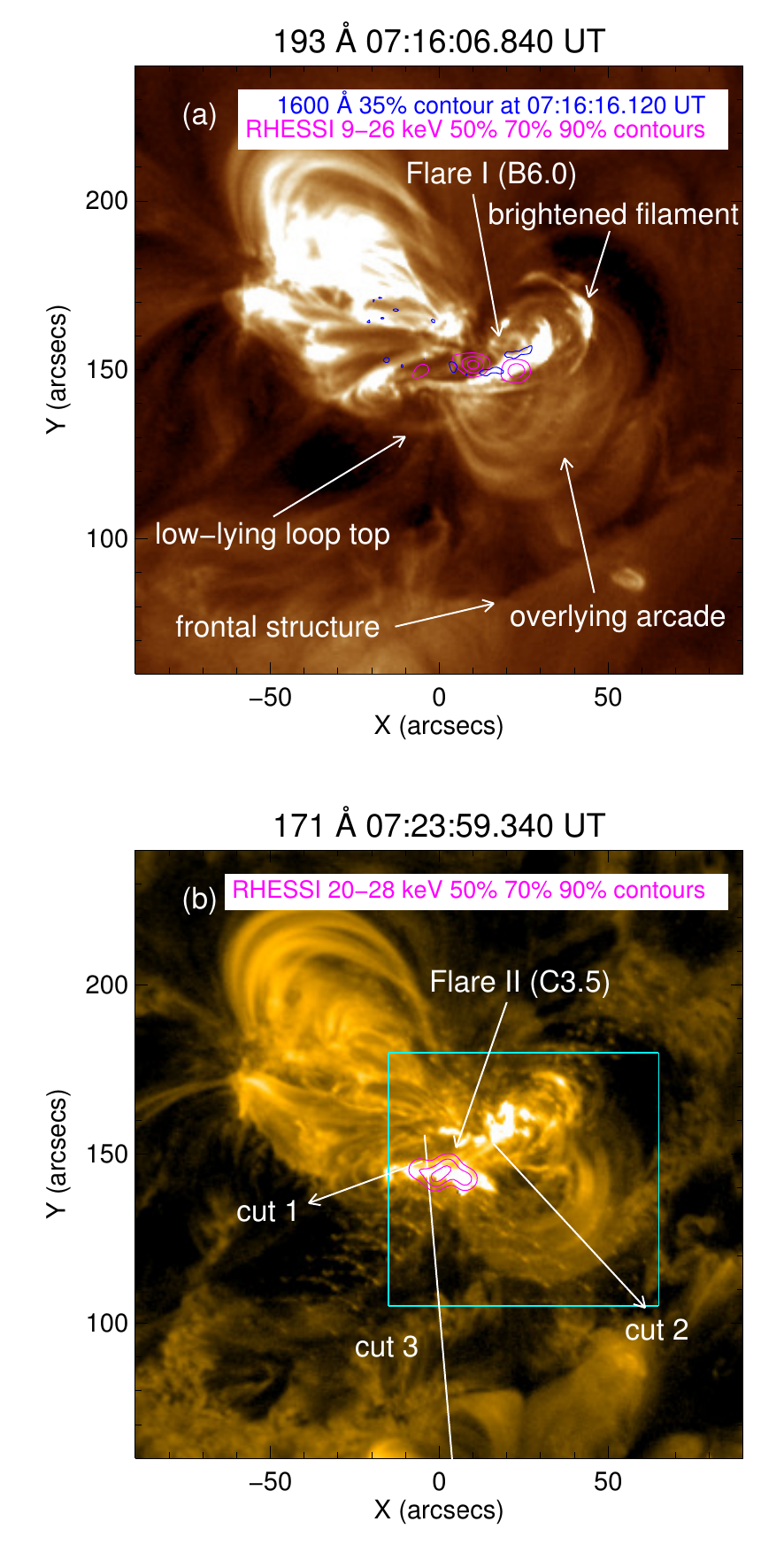}
\caption{\label{structure}Main features and processes identified in the 193 and 171 {\AA} passbands. (a) 193 {\AA}. The magenta contours for the 9-26 keV RHESSI HXR emission are integrated from 07:15:00 UT to 07:16:12 UT. (b) 171 {\AA}. The magenta contours for the 20-28 keV RHESSI HXR emission are integrated from 07:22:00 UT to 07:24:36 UT. Cut 1 is used to make the timeslices for the filament's eastern part in Figure~\ref{fevolve}(a) (the filament's eastern and western parts are denoted in Figure~\ref{filament}(a) with the same FOV); cut 2 for the filament's western part and the overlying arcade in Figure~\ref{fevolve}(b); cut 3 for the low-lying loop top and the frontal structure in Figure~\ref{fevolve}(c). The arrowhead of cut 3 is beyond the image edge. The cyan rectangular region is used to make AIA lightcurves in Figure~\ref{fevolve}(f). A color version of this figure is available in the online journal.}
\end{figure}

As the AR evolves, different features are identified, and we select two images in Figure~\ref{structure} to illustrate. As can be seen, a bright arcade overlies a sheared filament in the core region, and in the south, there is a curved frontal structure. These three features are visible clearly before the flare-associated evolution of the AR. Flare I, Flare II and low-lying loops in Figure~\ref{structure} appear during the following flare evolution, which will be discussed in the sections below. In the northeast, there is a large J-shaped arcade and some complex features underlying, but they are not involved in the activity in an apparent way, and thus will not be studied. 

The main C3.5 flare was preceded by a microflare B6.0 \citep[for a review of microflares see][]{han2011}, also associated with this AR, as evidenced by the AIA 1600~{\AA} ribbons and HXR emission imaged by RHESSI at 9-26 keV, as shown in Figure~\ref{structure}(a). Hereafter, we call this B6.0 microflare Flare I, and the subsequent C3.5 flare in Figure~\ref{structure}(b) Flare II. 

From $\sim$ 07:10 UT to 07:30 UT, the above features produce a rich sequence of phenomena. Figure~\ref{filament} illustrates the dynamical evolution of the filament in 304 {\AA}. The filament positions obtained from Figure~\ref{filament} are then overlaid on the contemporary 171 {\AA} images in Figure~\ref{arcfila}, which allows us to simultaneously track the evolution of the filament and the overlying arcade. Here we use the informative Figure~\ref{arcfila} to briefly summarise the main phenomena and their evolution, which are also listed in Table~\ref{timing}. More detailed information about the evolution will be described in the following subsections. Firstly, in Figure~\ref{arcfila}(a), the filament located near the site of Flare I is disrupted at the time when Flare I peaks ($\sim$ 07:15:40 UT). It then brightens and starts to distort, and a bump or bend in the filament moves from west to east (Figure~\ref{arcfila}(b) to (d)). This appears to push the overlying arcade upward and aside. When most of the bump suddenly escapes from beneath the overlying arcade ($\sim$ 07:22 UT, Figure~\ref{arcfila}(e)), the filament's eastern part erupts and the overlying arcade starts to contract. Almost at the same time, Flare II happens. From Figure~\ref{arcfila}(f) to (h), the inner loops of the overlying arcade continue contracting until when the GOES 1-8 {\AA} derivative reaches its peak ($\sim$ 07:25:45 UT; GOES lightcurves can be derived later in Figure~\ref{fevolve}(e)). Finally, the entire overlying arcade disappears in AIA 171 {\AA} (Figure~\ref{arcfila}(i)). Figure~\ref{c4aia} combines different wave bands (171, 211, 304 and 94 {\AA}) to illustrate the main events happening during the impulsive phase of Flare II for readers' convenience.

In Figure~\ref{structure}(b), we select three cuts to make timeslices for demonstrating the dynamical evolution of the filament's eastern part (cut 1), the overlying arcade and the filament's western part (cut 2), and the frontal structure and the low-lying loops (cut 3). The obtained timeslices, along with RHESSI HXR, GOES soft X-ray (SXR) and AIA lightcurves are collected in Figure~\ref{fevolve}. In the subsections below, combined with the information in Figure~\ref{fevolve}, we describe the processes in detail in order to give readers a more complete picture of the event.

\subsection{Flare I and Filament Eruption}\label{eruption}
\begin{figure*}
\includegraphics[scale=0.99]{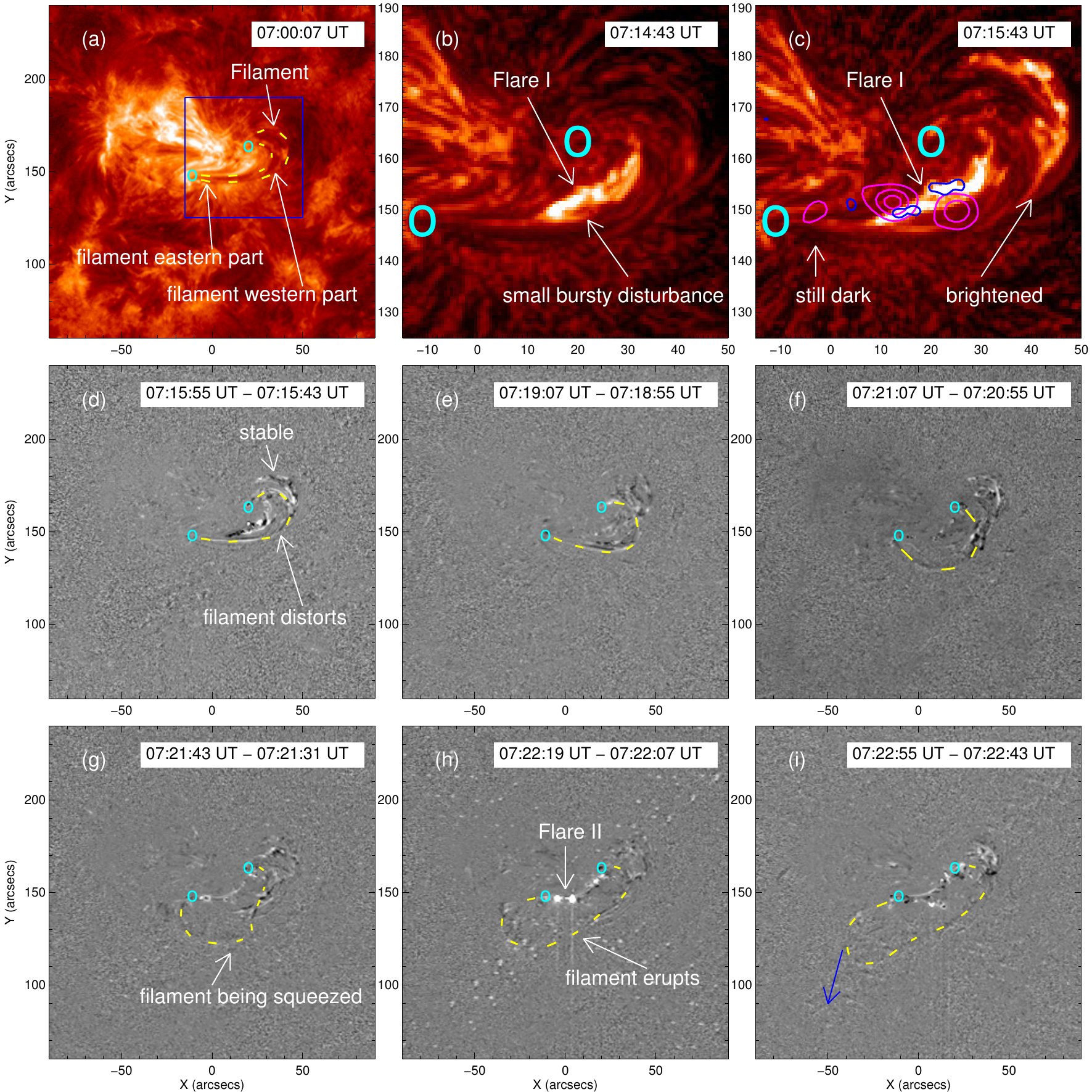}
\caption{\label{filament}Dynamical evolution of the filament in 304 {\AA}. (a) Denotes the position of the filament. The two cyan circles indicate the rough locations of the footpoints of the filament. The blue square region is the FOV used in (b) and (c). (b) Zoom in to show Flare I and its produced small bursty disturbance. (c) Zoom in to show the brightening filament's western part and still dark eastern part around the peak of Flare I. The blue contours are in 1600 {\AA} at $\sim$ 07:15:52 UT. The magenta contours are the same as in Figure~\ref{structure}(a). (d)-(i) Running difference images show the subsequent distortion and eruption of the filament. The yellow dashed line represent the shape and position of the filament in each image. The filament after (i) is too weak to be located, but still can be seen in the 304 {\AA} animation in this figure because it is moving. The blue arrow in (i) indicates the erupting direction of the filament's eastern part, and its head points to the rou
gh location of the filament top when it disappears. It should be noted that (b) and (c) have been processed using the multi-scale Gaussian normalisation (MGN) procedure \citep{mor2014}; the (d)-(i) running difference images are created after being processed using the MGN procedure. A color version of this figure is available in the online journal. An animation can be downloaded at the link \url{http://dx.doi.org/10.5525/gla.researchdata.363}, which is for this arXiv preprint version.}
\end{figure*}

\begin{figure*}
\includegraphics[scale=0.99]{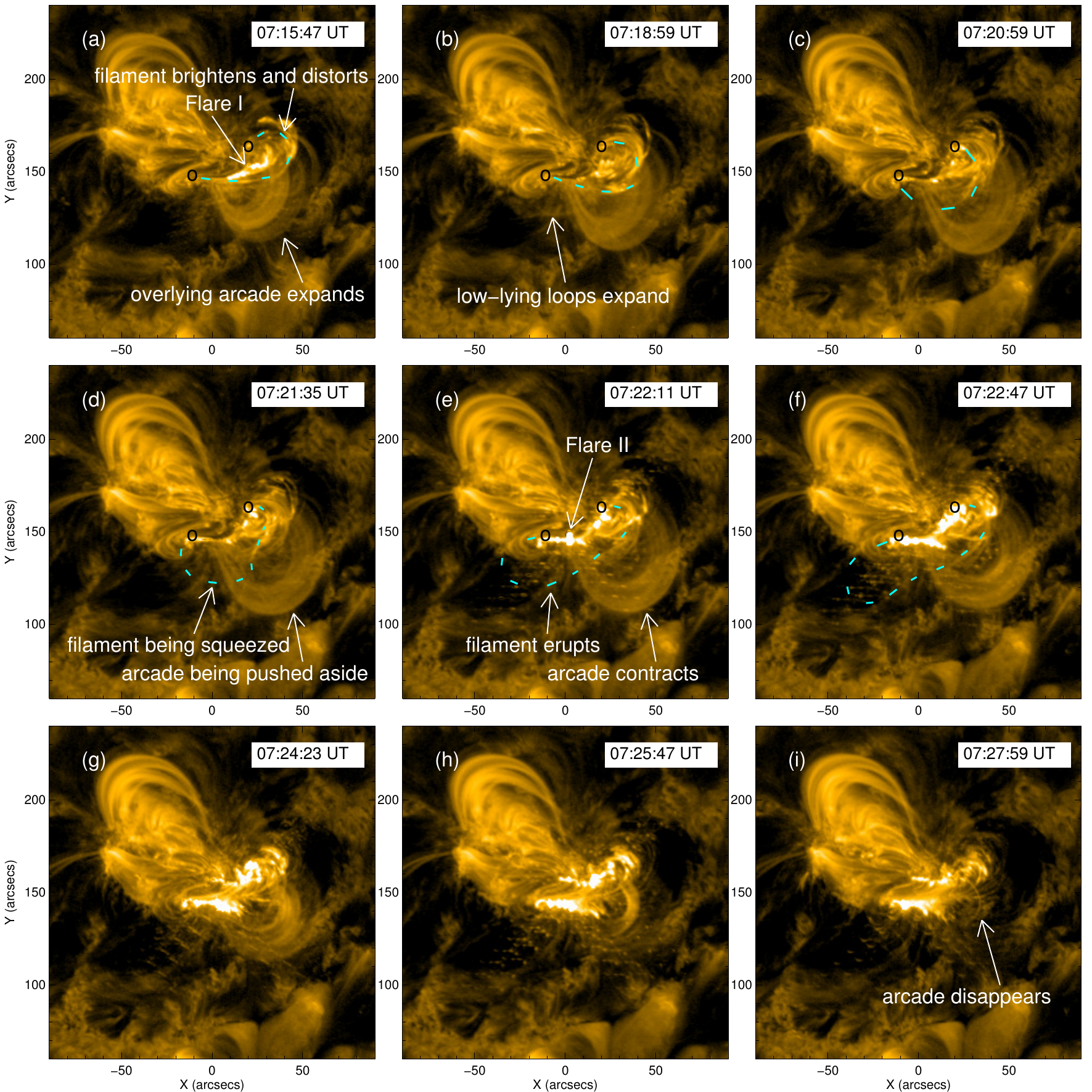}
\caption{\label{arcfila}Dynamical evolution of the overlying arcade in 171 {\AA}. The contemporary position of the filament obtained via the 304 {\AA} running difference image as in Figure~\ref{filament} is overlaid in each image, if possible. The filament after (f) is too weak to be located, but still can be seen in the 304 {\AA} animation in Figure~\ref{filament} because it is moving. A color version of this figure is available in the online journal. An animation can be downloaded at the link \url{http://dx.doi.org/10.5525/gla.researchdata.363}, which is for this arXiv preprint version.}
\end{figure*}

\begin{figure*}
\begin{centering}
\includegraphics[scale=0.9]{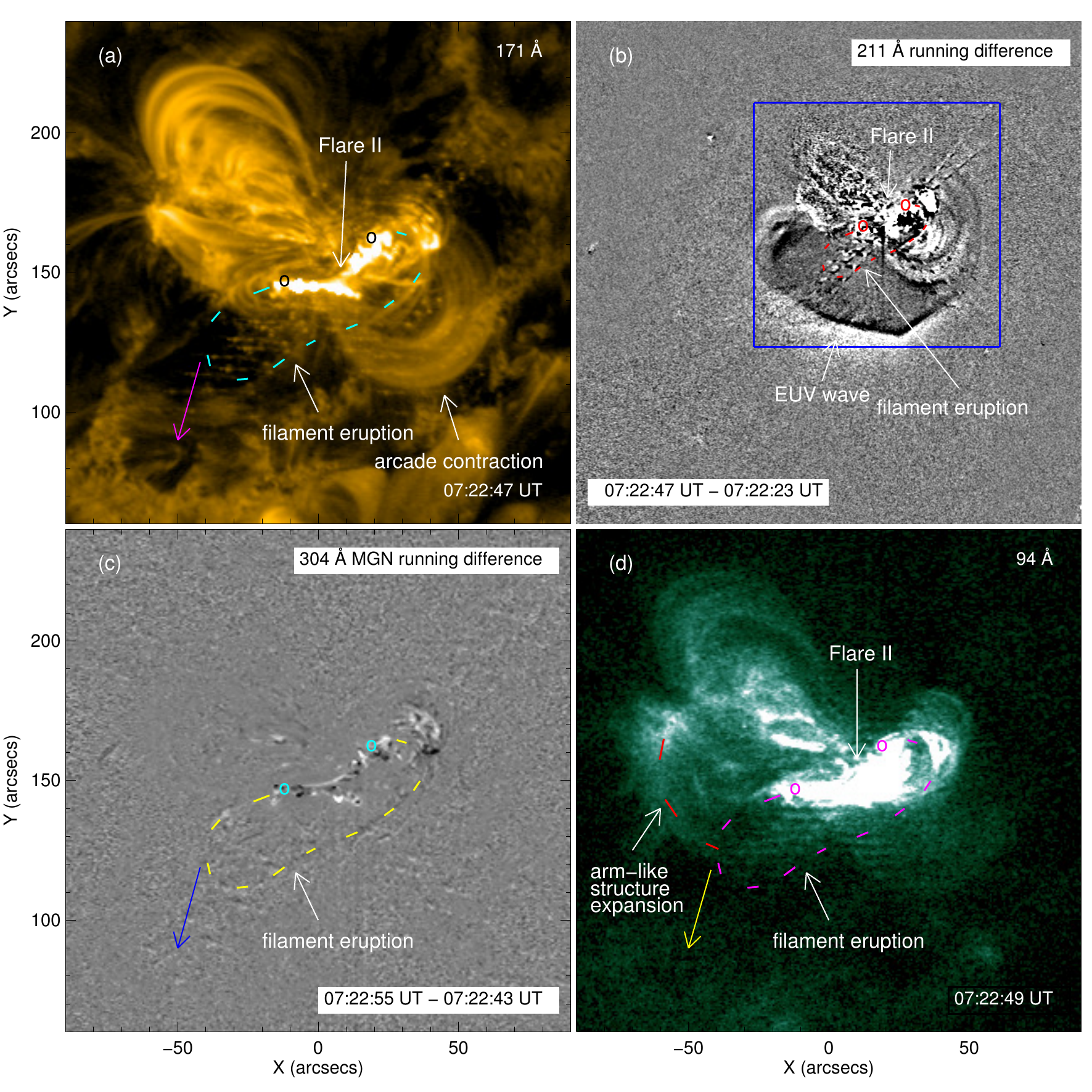}
\caption{\label{c4aia} Different wave bands show the main events simultaneously during the impulsive phase of Flare II. (a) Arcade contraction in 171 {\AA}, same as Figure~\ref{arcfila}(f). (b)  EUV wave showed by 211 {\AA} running difference. The blue square region is the FOV of the other three. (c) Filament eruption showed by 304 {\AA} MGN running difference, same as Figure~\ref{filament}(i). (d) Arm-like structure expansion in 94 {\AA}, same as Figure~\ref{armstruc}(h). The contemporary position of the filament obtained via the 304 {\AA} running difference image as in Figure~\ref{filament} is overlaid in each image. The arrow located around (-50, 100) in (a), (c) and (d) indicates the erupting direction of the filament's eastern part, and its head points to the rough location of the filament top when it disappears. A color version of this figure is available in the online journal. An animation can be downloaded at the link \url{http://dx.doi.org/10.5525/gla.researchdata.363}, which is for this arXiv preprint version.}
\end{centering}
\end{figure*}

\begin{figure*}[]
\includegraphics[scale=0.99]{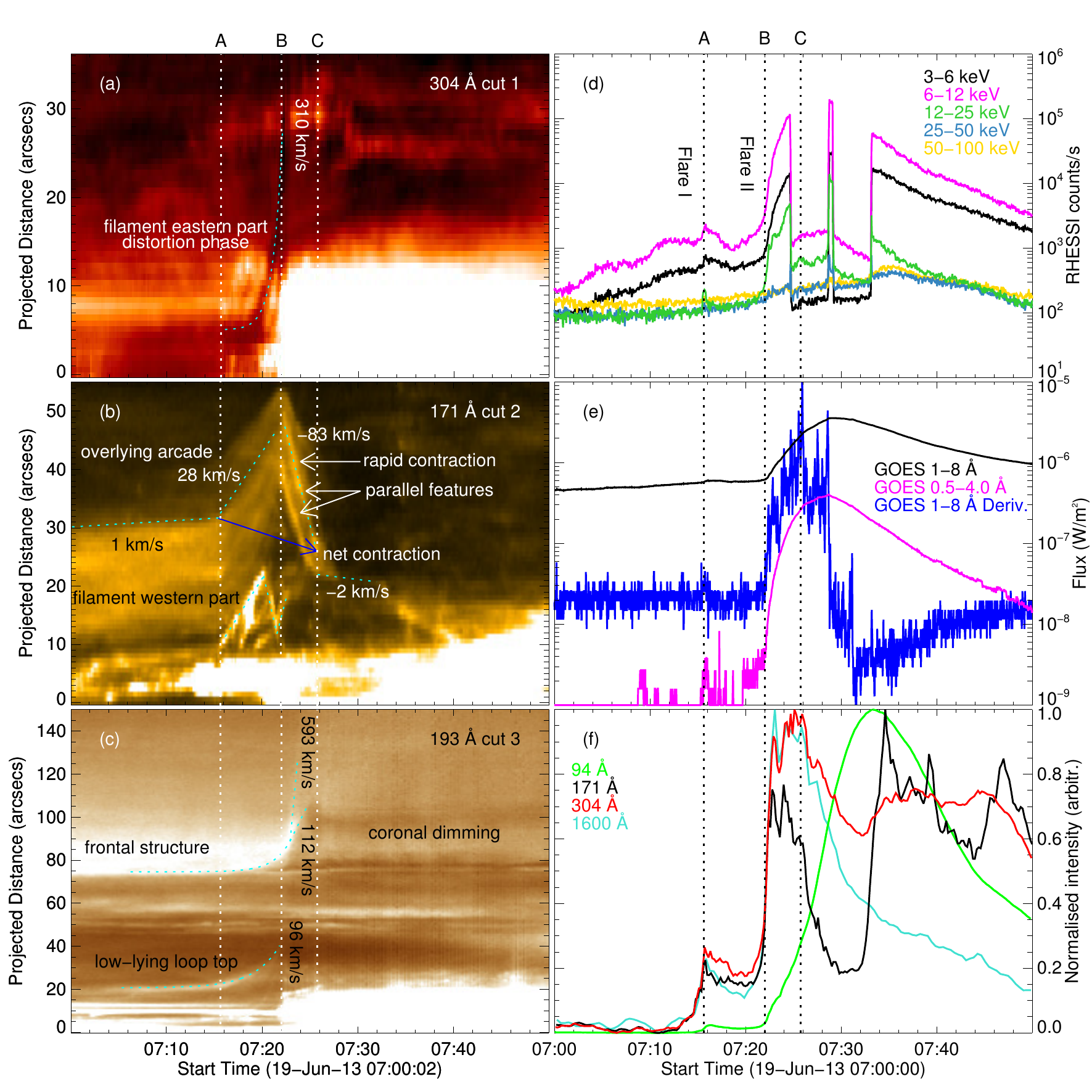}
\caption{\label{fevolve}Evolution of the flare. The cuts for the timeslices in (a)-(c) are shown in Figure~\ref{structure}(b). Different wavebands are used for cut 1, 2 and 3, because some features studied can only be clearly seen in specific wavebands. The letters A, B and C above the figures are used to denote the the main event timings in Table~\ref{timing}. (a) The timeslices in 304 {\AA} for cut 1 only show the distortion phase of the filament's dark eastern part. Its following dramatic eruption after 07:22 UT, unfortunately, cannot be tracked, because it is too weak (see the text in Section~\ref{eruption} for detailed explanation), but it still can be seen in the 304 {\AA} animation in Figure~\ref{filament} because it is moving. (b) The timeslices in 171 {\AA} for cut 2 show the expansion and contraction of both the overlying arcade and the filament's western part. (c)  The timeslices in 193 {\AA} for cut 3 show the expansion of both the frontal structure and the top of the low-lying loops. (d)
 RHESSI lightcurves in different wave bands. Note that the gradual increases at 3-6 keV and 6-15 keV from 07:00 UT until the small bump around Flare I are contributed by a limb event rather than this AR considered here. (e) GOES lightcurves. The GOES 1-8 {\AA} derivative has been normalised to fit the panel. (f) Normalised AIA lightcurves within the cyan rectangular region of Figure~\ref{structure}(b). A color version of this figure is available in the online journal.}
\end{figure*}

After analysing RHESSI images, we note that the gradual increase at 3-6 keV and 6-12 keV from 07:00 UT to about 07:15 UT in Figure~\ref{fevolve}(d) is contributed by a limb event (no HXR source can be detected). Only the small bump around 07:15:40 UT (indicated by ``A'' in Figure~\ref{fevolve}(d)) is the Flare I considered here, most prominent at RHESSI 12-25 keV and AIA lightcurves in Figure~\ref{fevolve}(d) and (f), respectively. Its two ribbons in AIA 1600 {\AA} and RHESSI HXR contours can be clearly seen in Figure~\ref{structure}(a) and Figure~\ref{filament}(c), just encircled by the nearby filament.  

Figure~\ref{filament} presents the activities in 304 {\AA}. At $\sim$ 07:00 UT, the filament has a sheared appearance, with its bump pointing to the west (Figure~\ref{filament}(a)). Then Flare I occurs, and at $\sim$ 07:14:43 UT, it seems to  produce a small bursty disturbance, pointing to the filament's western part (Figure~\ref{filament}(b)). Around 1 min later, at 07:15:43 UT when Flare I peaks (revealed by the RHESSI 12-25 keV lightcurve and indicated by ``A'' in Figure~\ref{fevolve}(d)), the filament's western part suddenly brightens, with some plasma flowing to its northern footpoint (seen in the 304 {\AA} animation in Figure~\ref{filament}), though the eastern part is still dark (Figure~\ref{filament}(c)). Subsequently, the filament becomes distorted, with its bump propagating from west to east, though there is still part of the filament remaining relatively stable (Figure~\ref{filament}(d)-(f)). The dark trajectory and the bright path denoted by ``filament eastern part'' and ``filament western part''  in Figure~\ref{fevolve}(a) and (b) just show the filament's eastern and western parts sweeping across cut 1 and 2 of Figure~\ref{structure}(b) during the distortion, respectively (an exponential line is overlaid in Figure~\ref{fevolve}(a) to approximate the trajectory). When the bump propagates close to the filament's eastern end, the western part contracts, which appears squeezed and highly energised, and the entire filament expands more outward (Figure~\ref{filament}(g)). Then in Figure~\ref{filament}(h) the eastern part erupts dramatically and nonradially, as a cool, extending feature at $\sim$ 07:22 UT (see the 304 {\AA} animation in Figure~\ref{filament}), and almost simultaneously, Flare II happens (indicated by ``B'' in Figure~\ref{fevolve}(d)). Such an eruption can be categorised as a whipping-like asymmetric filament eruption \citep{liu2009a,jos2013}. Because during the eruption the filament is too weak and vague, even in the running difference images, and Flare II produces strong 
 flashes,  we are not be able to select a cut to describe the following movement of the filament after $\sim$ 07:22 UT. Thus the trajectory in Figure~\ref{fevolve}(a) for cut 1 mainly demonstrates the kinematics of the filament's eastern part in the previous distortion phase before $\sim$ 07:22 UT, but the 304 {\AA} animation in Figure~\ref{filament} (and also its running difference version) can be taken as a reference for the following eruption of the filament's eastern part because of its moving nature.  The bright path denoted by ``filament western part''  in Figure~\ref{fevolve}(b) shows that the filament's western part expands again after $\sim$ 07:22 UT when the filament's eastern part erupts. The entire filament in Figure~\ref{filament}(h) seems relaxed from the squeezed state in Figure~\ref{filament}(g), like an elastic tube which can be stretched. The arrow in Figure~\ref{filament}(i) denotes the erupting direction of the filament's eastern part, and its head indicates the rough location of 
 the filament top when it disappears.

\begin{figure*}[]
\includegraphics[scale=0.99]{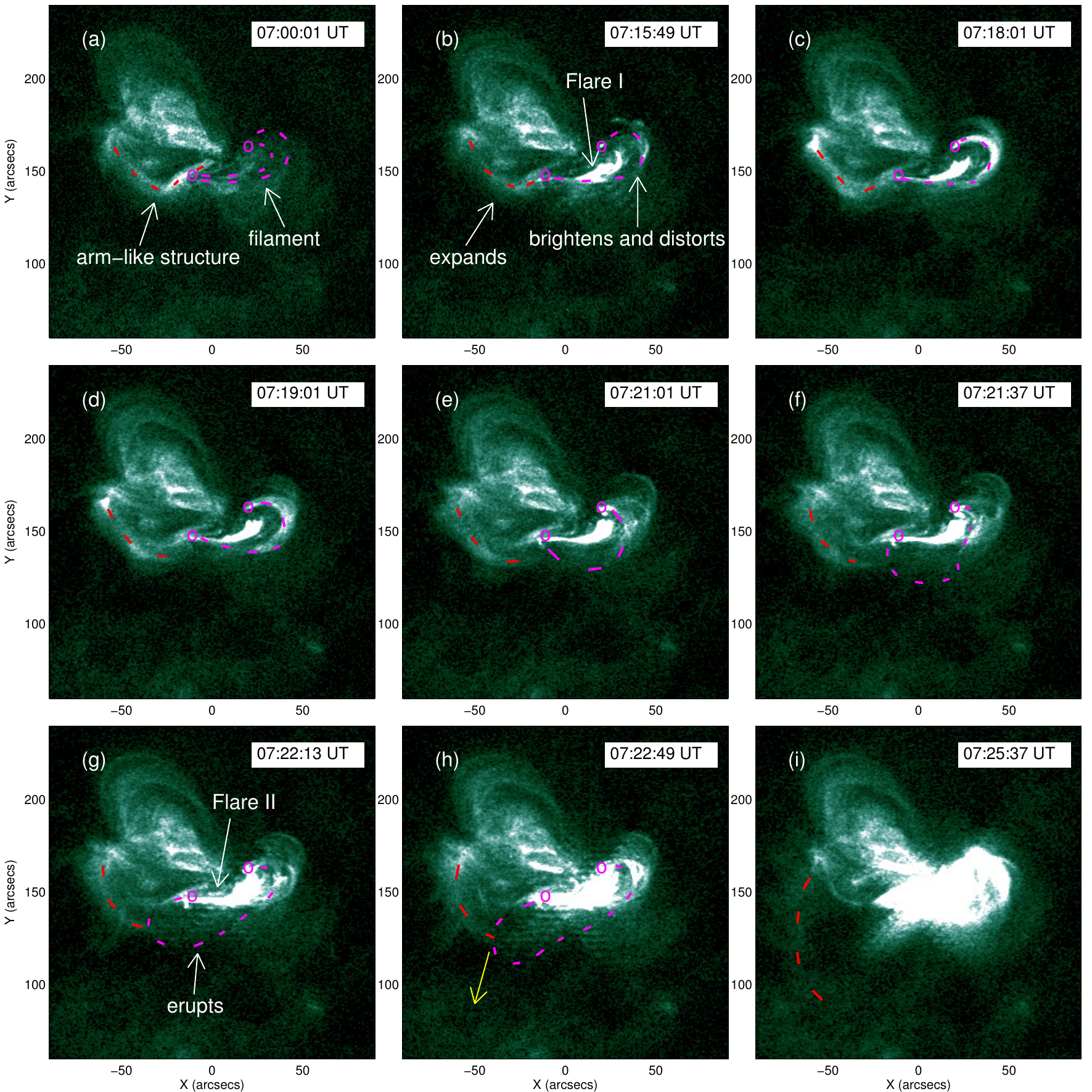}
\caption{\label{armstruc}Evolution of the arm-like structure in 94 {\AA}. The contemporary position of the filament obtained via the 304 {\AA} running difference image as in Figure~\ref{filament} is overlaid in each image, if possible. The yellow arrow in (h) indicates the erupting direction of the filament's eastern part, and its head points to the rough location of the filament top when it disappears, as in Figure~\ref{filament}(i). An animation of the 94 {\AA} evolution can be found below Figure~\ref{c4aia}. A color version of this figure is available in the online journal.}
\end{figure*}

\begin{deluxetable*}{cc}
\tabletypesize{\small}
\tablewidth{0pt}
 \tablecaption{The main evolution in SOL2013-06-19.\label{timing}}
 \tablehead{
 \colhead{Time}  & \colhead{Events}}
 \startdata
$\sim$ 07:15:40 (A) & Flare I peaks; \\
 & filament's western part brightens and starts to distort; \\
 & overlying arcade starts to expand. \\
 \hline
$\sim$ 07:22:00 (B) & filament's eastern part erupts; \\
& Flare II starts; \\
& overlying arcade starts to contract. \\
 \hline
$\sim$ 07:25:45 (C) & inner loops of the overlying arcade contract to a relatively stable position; \\
& GOES 1-8 {\AA} derivative reaches its peak.
\enddata
\tablenotetext{a}{The letters A-C are used in Figure~\ref{fevolve} to indicate the event timings.}
\end{deluxetable*}

\subsubsection{Other Structures Associated with the Filament Eruption}\label{exstruc}
An interesting development is that an eastern arm-like structure also brightens and expands outwards with the filament (see Figure~\ref{c4aia}(d) and Figure~\ref{armstruc}), which can only be clearly seen in 94 {\AA}. We again overlay the positions of the contemporary filament obtained from the 304 {\AA} running difference images like in Figure~\ref{filament} onto the 94 {\AA} images in Figure~\ref{c4aia}(d) and Figure~\ref{armstruc}. The final projected positions of the expanding portion of this arm-like structure and of the filament's erupting top seem near to each other before they disappear, around the southeastern corner of Figure~\ref{armstruc}(i) (the arrows in Figure~\ref{c4aia}(d) and Figure~\ref{armstruc}(h) denote the erupting direction of the filament's eastern part, and its head indicates the rough location of the filament top when it disappears, like in Figure~\ref{filament}(i)).  This might make a reconnection between the arm-like structure and the filament possible, which will be discussed in Section~\ref{reconnection}.

In Figure~\ref{structure}(a), far to the south of the filament, there is a
frontal structure, most prominent in 193 and 211 {\AA}. It exists even before the two flares, and could be a stable cavity edge as described in \citet{hud1999}.
This global structure is similar to that of a CME with a filament at the bottom, a cavity in the middle and a frontal loop at the top. From the timeslices in Figure~\ref{fevolve}(c) for cut 3 of Figure~\ref{structure}(b), it can be seen that the frontal structure also starts expanding
exponentially at $\sim$ 07:15:40 UT when Flare I peaks. At $\sim$
07:22:40 UT, it begins to diffuse with a leading edge $\sim593$
km/s and a trailing edge  $\sim112$ km/s. And behind the trailing edge, a coronal dimming appears. This may be consistent with the hybrid EUV wave model, with a fast-mode wave component ahead of a CME-driven compression front \citep[see][and references therein]{wei2014}.  In addition, in the 211 {\AA} running difference animation in Figure~\ref{c4aia}, we also note that there are quasi-periodic wave trains accompanying the EUV wave \citep[see][and references therein]{wei2014}. Here we just point out that an EUV wave with quasi-periodic wave trains exists in this event, which is associated with the expanding frontal structure, and also suggest that it should be added in the list at the link \url{http://www.lmsal.com/nitta/movies/AIA_Waves/oindex.html} for future study. No further discussion will be presented because it is beyond the scope of this paper.

\subsection{Overlying Arcade Expansion \& Contraction}\label{contraction}
Figure~\ref{arcfila} illustrates the dynamical evolution of the overlying arcade, overlaid by the contemporary positions of the filament. The timeslices in Figure~\ref{fevolve}(b) for cut 2 of Figure~\ref{structure}(b) show that the overlying arcade has a small increase in
height from 07:00 UT to $\sim$ 07:15:40 UT. Then at $\sim$ 07:15:40 UT when Flare I peaks and the filament starts to distort, it accelerates to expand at a
nearly uniform apparent speed of $\sim28$ km/s (Figure~\ref{arcfila}(a) to (d)). In the 171 {\AA} animation in Figure~\ref{arcfila}, it also seems to be pushed aside and incline towards the solar disk during the end of this expansion phase (see also Figure~\ref{arcfila}(d)). The low-lying loops (Figure~\ref{structure}(a) and Figure~\ref{arcfila}(b)) overlying the filament's eastern part appear and also start to expand (revealed by the timeslices in Figure~\ref{fevolve}(c) for cut 3 of Figure~\ref{structure}(b)).  At $\sim$ 07:22:00 UT when the filament's eastern part erupts and Flare II occurs, the overlying arcade motion turns to a
rapid contraction at a nearly constant apparent speed of $\sim83$ km/s (Figure~\ref{arcfila}(e)). Figure~\ref{arcfila}(e) to (h) show that a moderate inclination of the arcade seems to accompany the rapid contraction (also see the 171 {\AA} animation in Figure~\ref{arcfila} after $\sim$ 07:22 UT). Shown in Figure~\ref{fevolve}(b), the inner loops of the arcade contract rapidly by about a half with respect to the starting position of cut 2 until $\sim$ 07:25:45 UT (the starting position of cut 2 is around the middle of the two footpoints of the contracting arcade, as can be seen in Figure~\ref{structure}(b)), which also can be seen by comparing Figure~\ref{arcfila}(e) with (h). As the rapid contraction of the inner loops stops, the derivative of GOES 1-8 {\AA} flux peaks (indicated by ``C'' in Figure~\ref{fevolve}(e)). Thus the rapid contraction may only happen during the rise stage of the impulsive phase (Neupert effect; \citeauthor{neu1968} \citeyear{neu1968}). The projected net contraction of the 
arcade indicated by the blue arrow (which connects the beginning of the rapid expansion to the ending of the rapid contraction) in Figure~\ref{fevolve}(b) is $\sim$ 4.5 arcsecs. At the end, the entire overlying arcade disappears (Figure~\ref{arcfila}(i)).

\section{Magnetic Field Extrapolation} \label{extrapolation}
We employ a nonlinear force free field (NLFFF) model approach in order to explore the coronal magnetic field configuration before and after the C3.2 flare. Photospheric vector magnetograms obtained by SDO/HMI between 06:00 UT and 09:00 UT (excluding the one at 07:24 UT when the violent C3.2 flare happens), with a 12 minute cadence (the vector data is also averaged in a 12-minute period) and a $\sim1.0$~arcsec spatial resolution, are used as input to our modeling. The extension of our model volume is $\approx331\times258\times129$~arcsec, i.e., $\approx244\times190\times95$~Mm, centered around solar $(x,y)=(-28.2,137.9)$~arcsec. This proximity of the considered area to the disk center allows us to neglect eventual projection (foreshortening) effects. The vertical magnetic flux within the area is balanced to within $\approx10\%$. 

Using standard IDL mapping software, we de-rotate the measured magnetic field vector maps to the flare peak time and project the data to a local coordinate system \citep[following][]{gar1990}. The observed non-force-free photospheric data is driven to a more force-free consistent field configuration, following \cite{wie2006}, which is then supplied to the NLFFF modeling scheme as a lower boundary condition (for details of the method see \citeauthor{wie2010} \citeyear{wie2010}; \citeauthor{wie2012} \citeyear{wie2012};  and Section 2.2.1 of \citeauthor{der2015} \citeyear{der2015}).

In order to quantify the goodness of the obtained NLFFF model solutions we use some of the metrics introduced in \cite{whe2000}. First, we test the success of recovering a force-free solution using the current-weighted average of the sine of the angle between the model magnetic field and the electric current density, where we find $\sigma_j$ on the order of $10^{-1}$ (note that for a perfectly force-free solution one would find $\sigma_j =0$). Second, we calculate a measure for the solenoidality of the model solution, in the form of the volume-averaged fractional flux, and find $\langle|f_i|\rangle$ on the order of $10^{-4}$ (for a perfectly solenoidal solution one would find $\langle|f_i|\rangle=0$). That indicates that our NLFFF models are force-free and solenoidal to a necessary degree in order to validly approximate the pre- and post-flare coronal magnetic field.

\subsection{Overlying Arcade Contraction}\label{contract}
\begin{figure*}
\hspace{0cm}\includegraphics[width=0.452\textwidth]{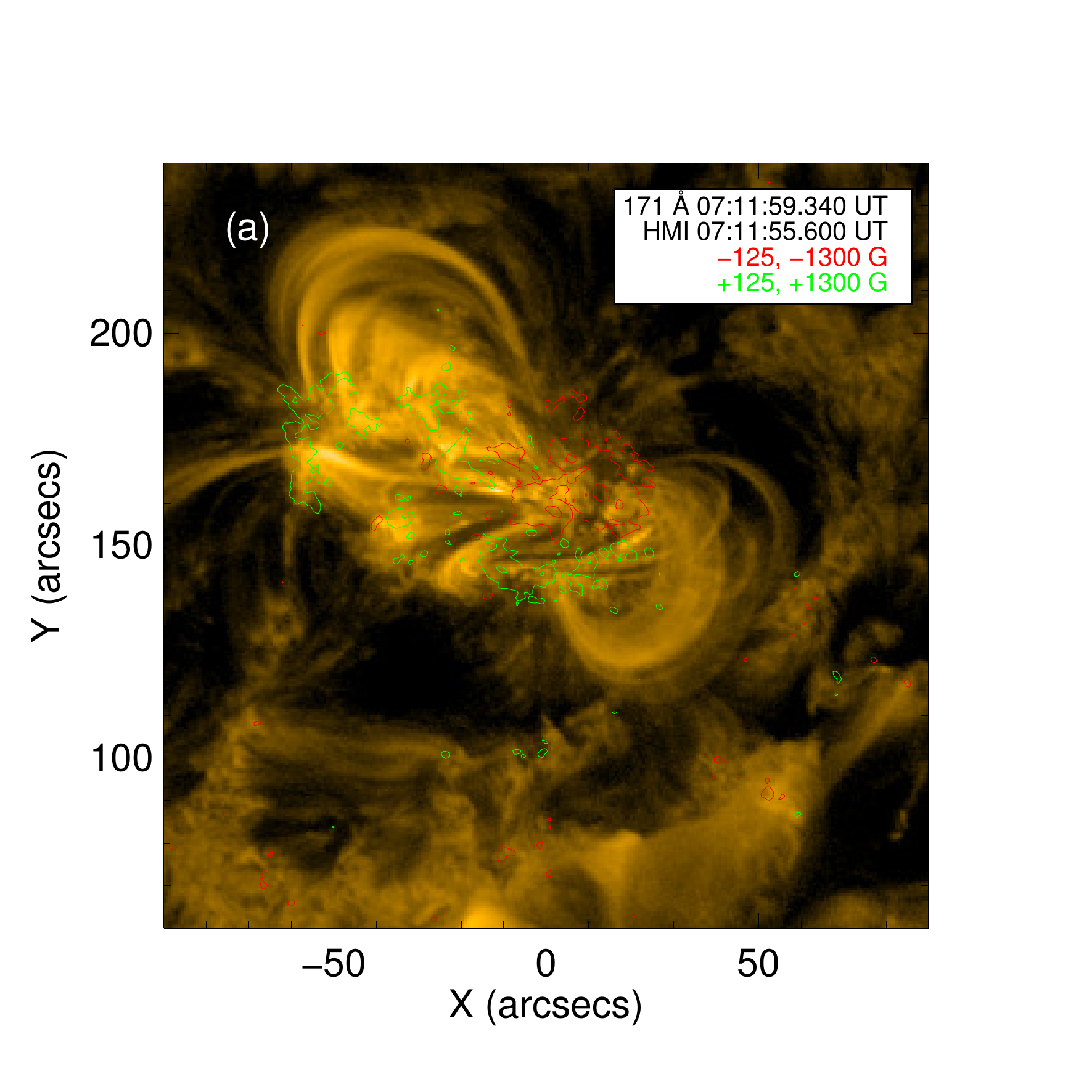}
\hspace{0cm}\includegraphics[width=0.452\textwidth]{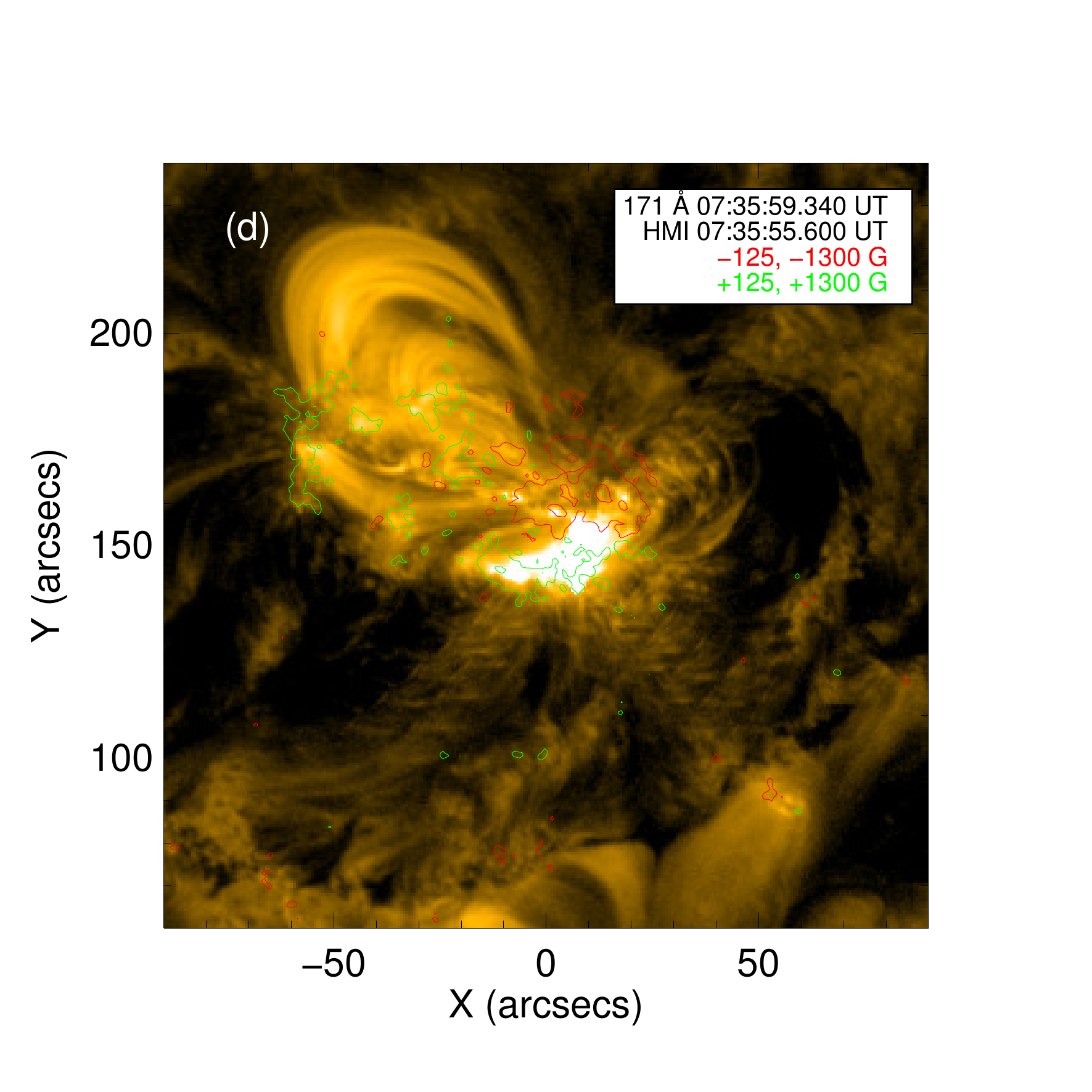}
\hspace{0cm}\includegraphics[width=0.452\textwidth]{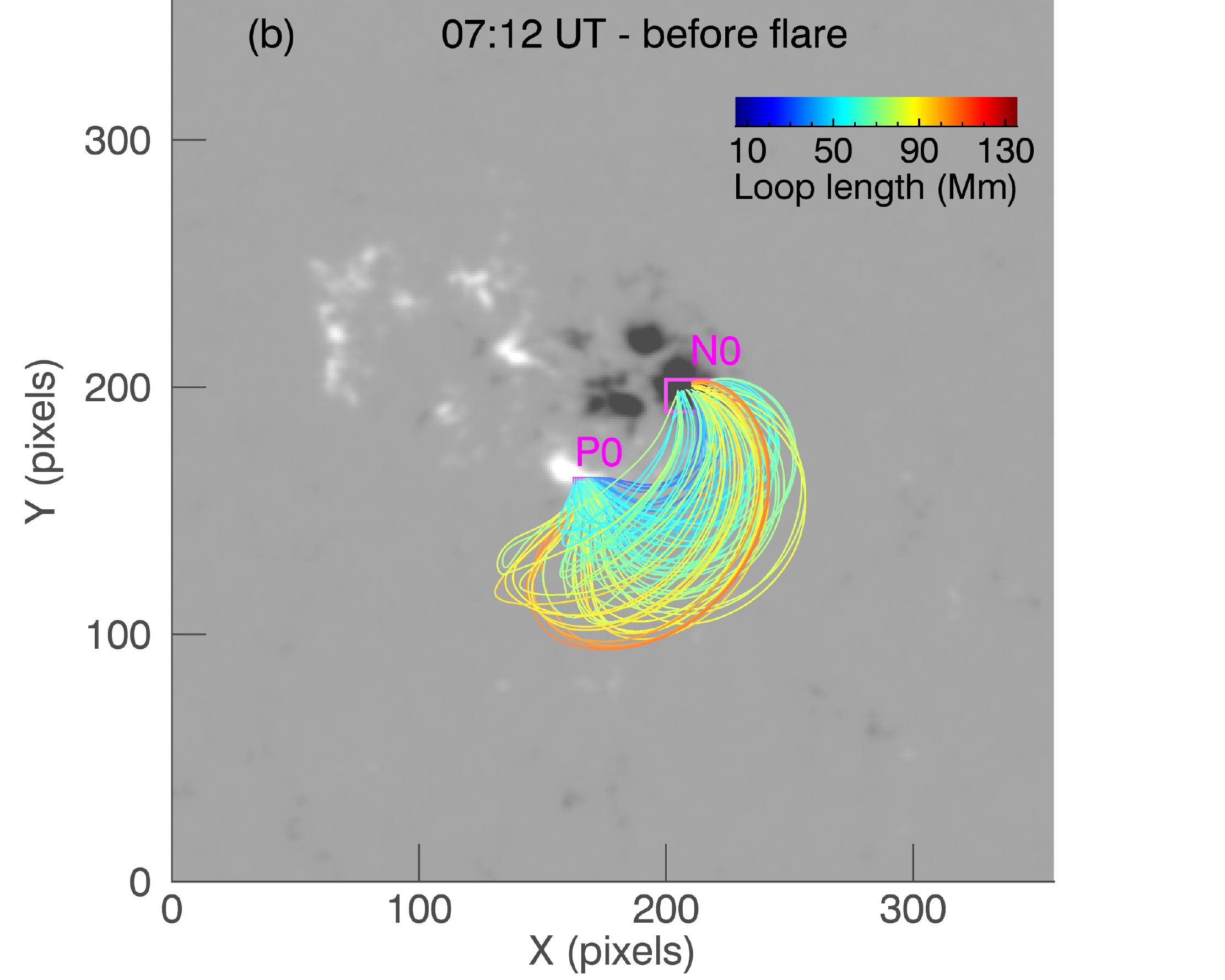}
\hspace{0cm}\includegraphics[width=0.452\textwidth]{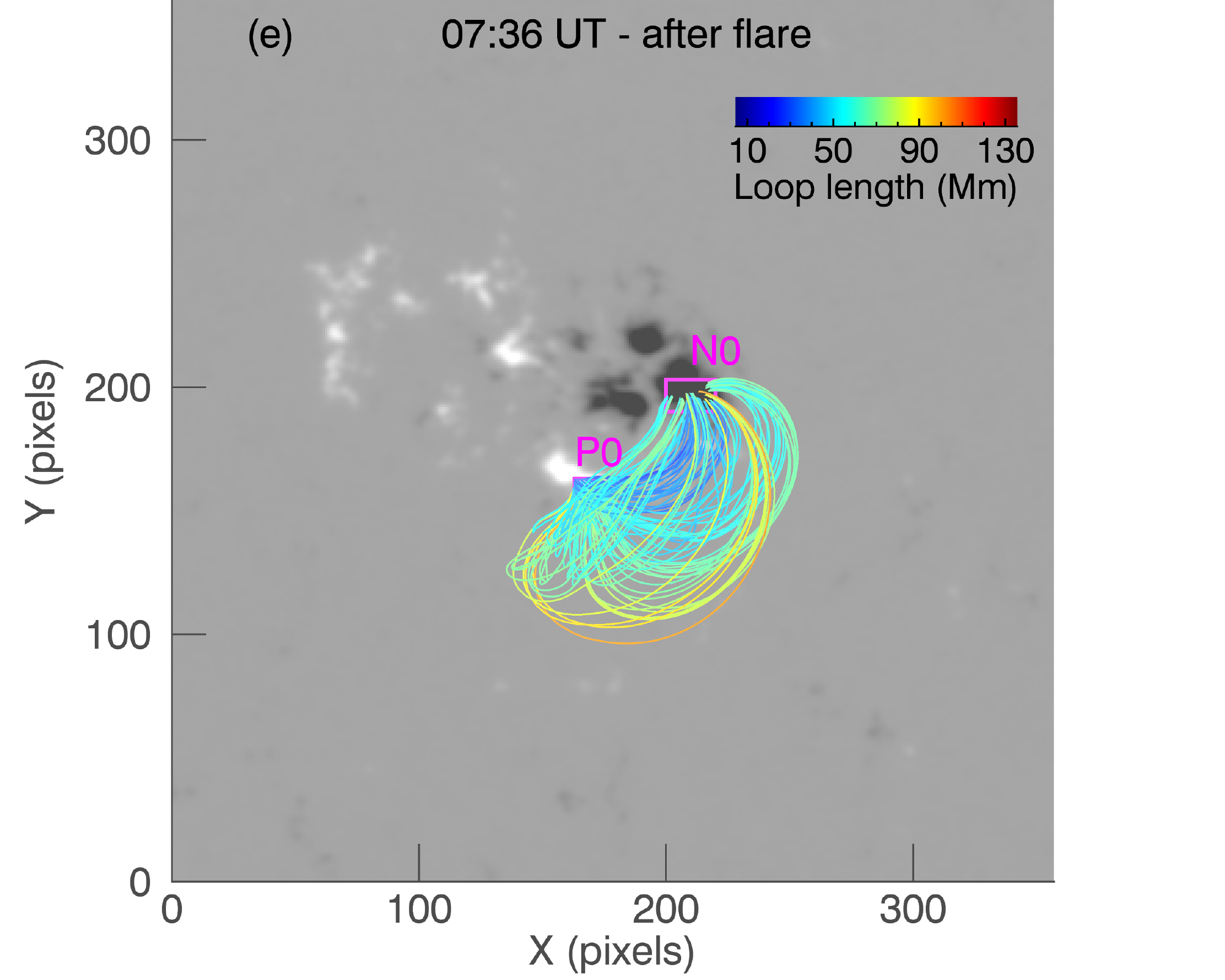}
\hspace{0cm}\includegraphics[width=0.452\textwidth]{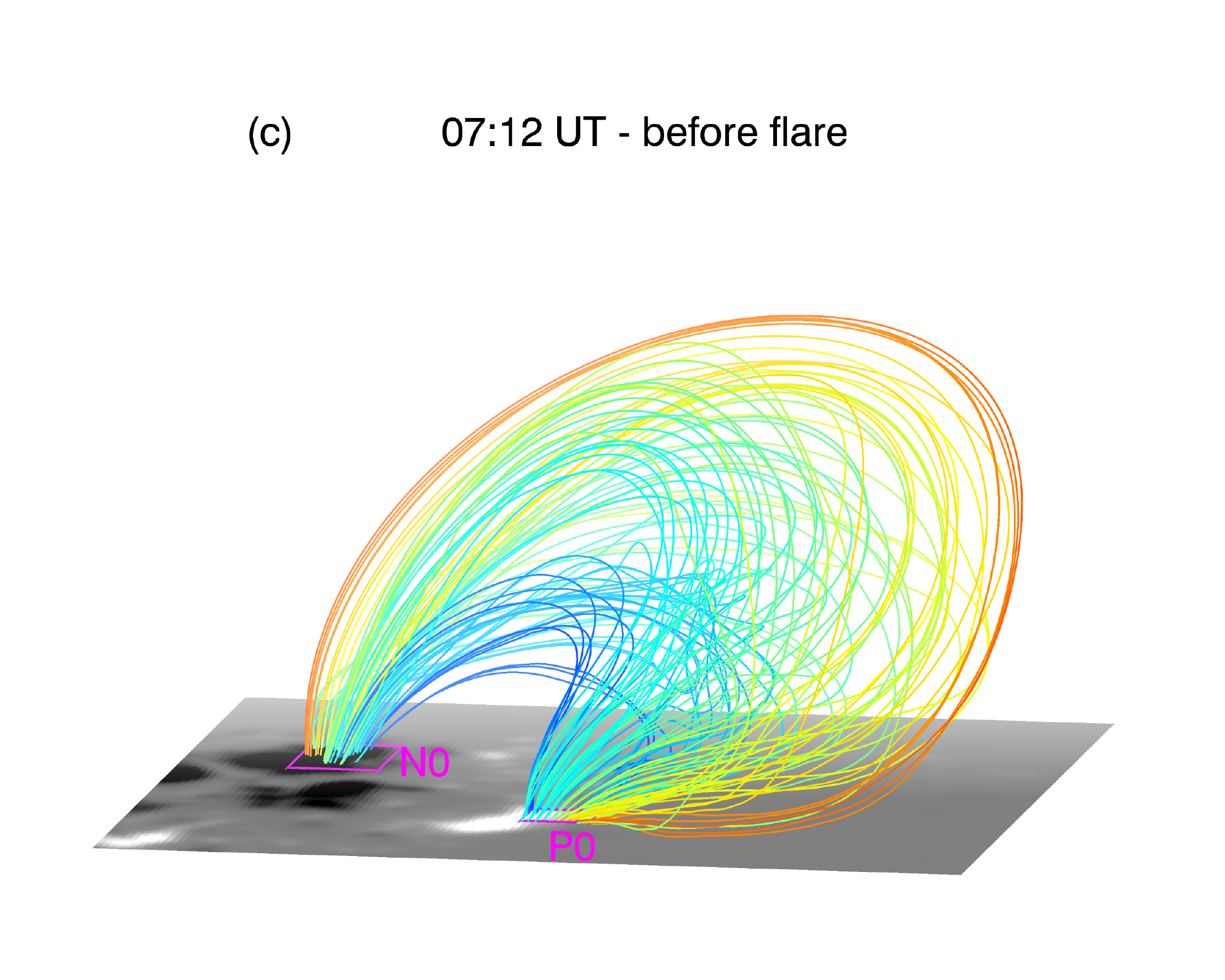}
\hspace{1.65cm}\includegraphics[width=0.46\textwidth]{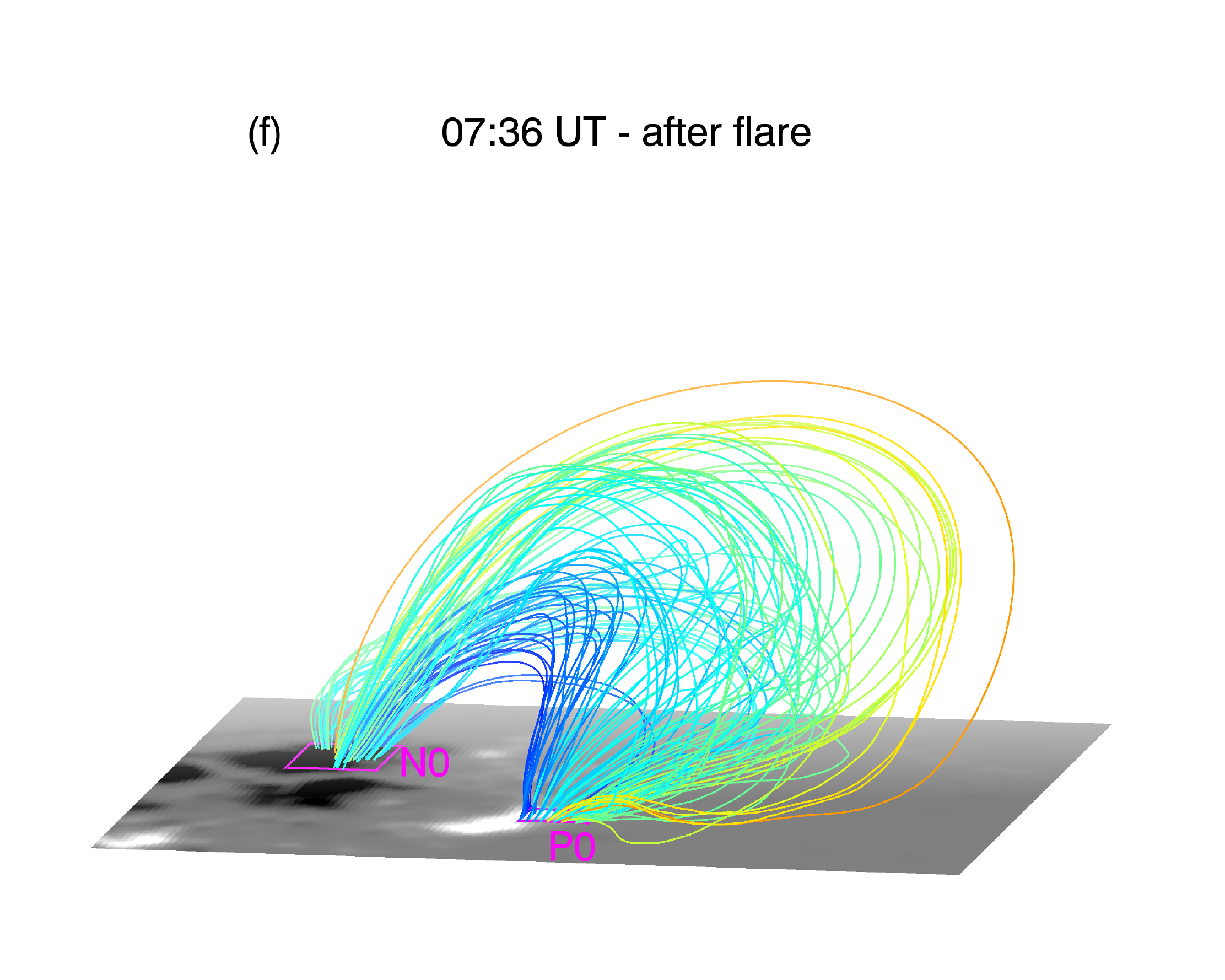}
\caption{\label{arcade}Overlying arcade contraction found in the extrapolation. (a) Longitudinal magnetogram overlaid on AIA 171 {\AA} image at around 07:12 UT (before flare) for comparison with extrapolation. (b) The overlying arcade in extrapolation at around 07:12 UT. P0 and N0 are the areas used to select the field lines. The FOV is approximately the same as in (a). 1 pixel $\approx$ 0.5 arcsecs. (c) 3D view of the overlying arcade at 07:12 UT. (d)-(f) Same as (a)-(c), but at 07:36 UT (after flare). A color version of this figure is available in the online journal.} 
\end{figure*}

\begin{figure}
\begin{centering}
\includegraphics[width=0.45\textwidth]{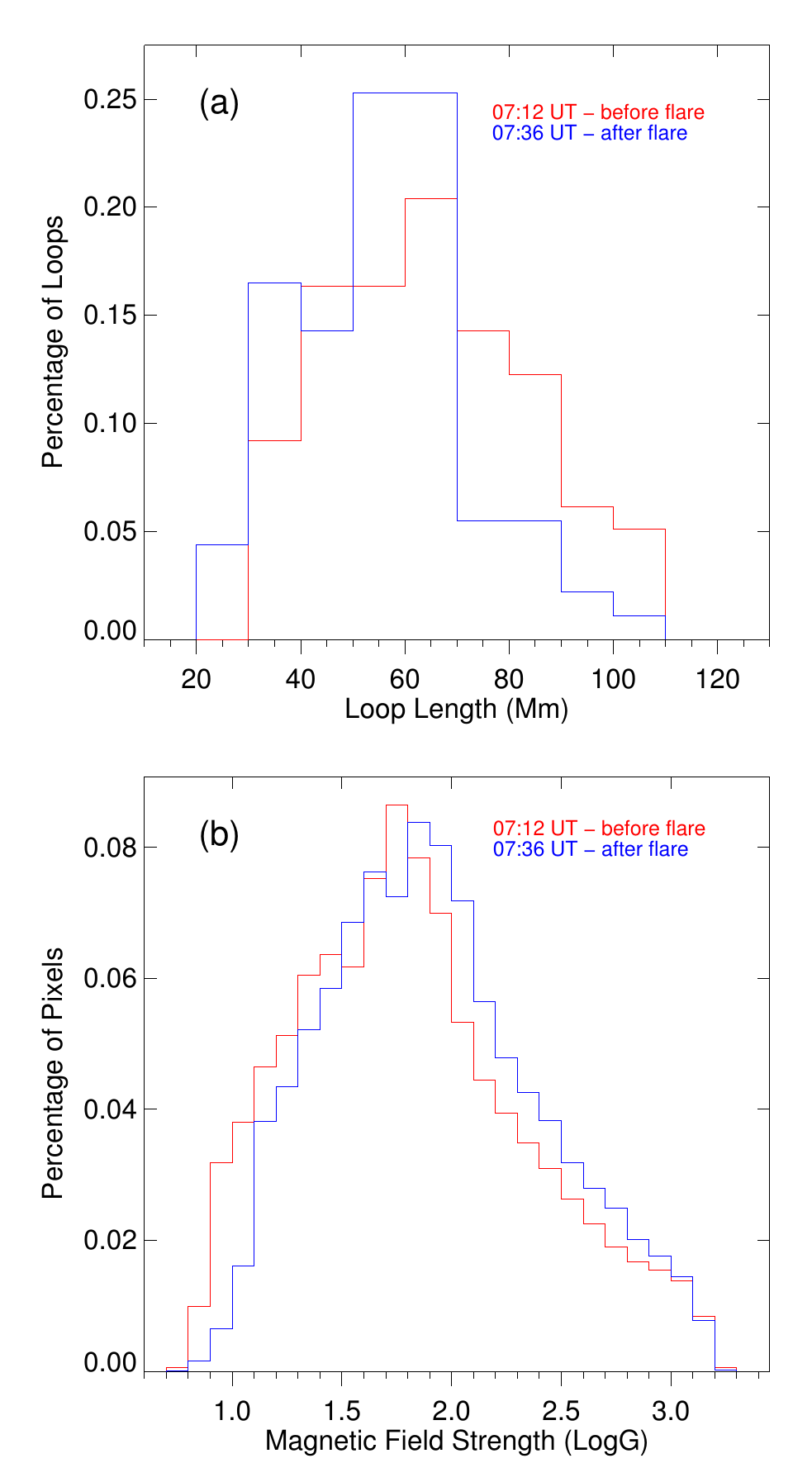}
\caption{\label{lengthhist}Changes in lengths and magnetic field strengths of the extrapolated overlying arcade field lines shown in Figure~\ref{arcade} between before and after flare. (a) Normalised histograms of the lengths of the arcade field lines. (b) Normalised histograms of the magnetic field strengths of all pixels of the extrapolated arcade. A color version of this figure is available in the online journal.} \end{centering}
\end{figure}

In order to picture the flare-associated magnetic field evolution, we trace model magnetic field lines from certain locations at the NLFFF model lower boundary. Since the photospheric field (used as input to the modelling) is evolving in time, the same coordinates at different times may correspond to physically different structures. Therefore, we use a group of field lines occupying a large regions, and study their statistics, which can diminish the above influence. We choose the area P0 (5 $\times$ 5 arcsecs, comparable to the overlying arcade footpoint area in the positive polarity region in AIA 171 {\AA}; see Figure~\ref{arcade}) as the leading footpoint region, that is the footpoint region from which the extrapolated field lines are calculated \citep{wie2013}. The area N0 defines the region where the arcade connects at the negative magnetic polarity. We take all the calculated field lines from P0 to N0 as the overlying arcade at different times. By visually comparing the 
arcades of considered model field lines between 07:12 UT and 07:36 UT in Figure~\ref{arcade}, it appears that the number of longer (red or yellow) field lines decreases and that of shorter (blue) ones increases (the total numbers of the field lines at these two times are comparable, thus the comparison is valid). This is more obvious from the normalised histograms of lengths of field lines in Figure~\ref{lengthhist}(a), with the fraction of longer field lines decreasing and that of shorter ones increasing after the flare. Globally, the histogram is shifted to shorter length. In addition, in Figure~\ref{lengthhist}(b), we construct normalised histograms of the field strengths at all pixels along all of the individual field lines in the reconstructed overlying arcade. They show that with the contraction, the magnetic field strength of the arcade is globally enhanced after the flare.

From AIA 171 {\AA} images it is not possible to detect the lower and shorter field lines in Figure~\ref{arcade}(b) and (e). Thus in order to compare the extrapolations with AIA observations, we choose the field lines with lengths larger than average, and calculate the average projected distances of the midpoints of the field lines to the midpoints of the lines connecting their conjugate footpoints at both 07:12 UT and 07:36 UT.  Their difference reflects the average projected contraction distance. The obtained value is $\sim4.7$ arcsecs, which is in good agreement with the net projected contraction $\sim$ 4.5 arcsecs observed in AIA 171 {\AA} (the blue arrow in Figure~\ref{fevolve}(b)). 

\begin{figure}
\begin{centering}
\includegraphics[width=0.45\textwidth]{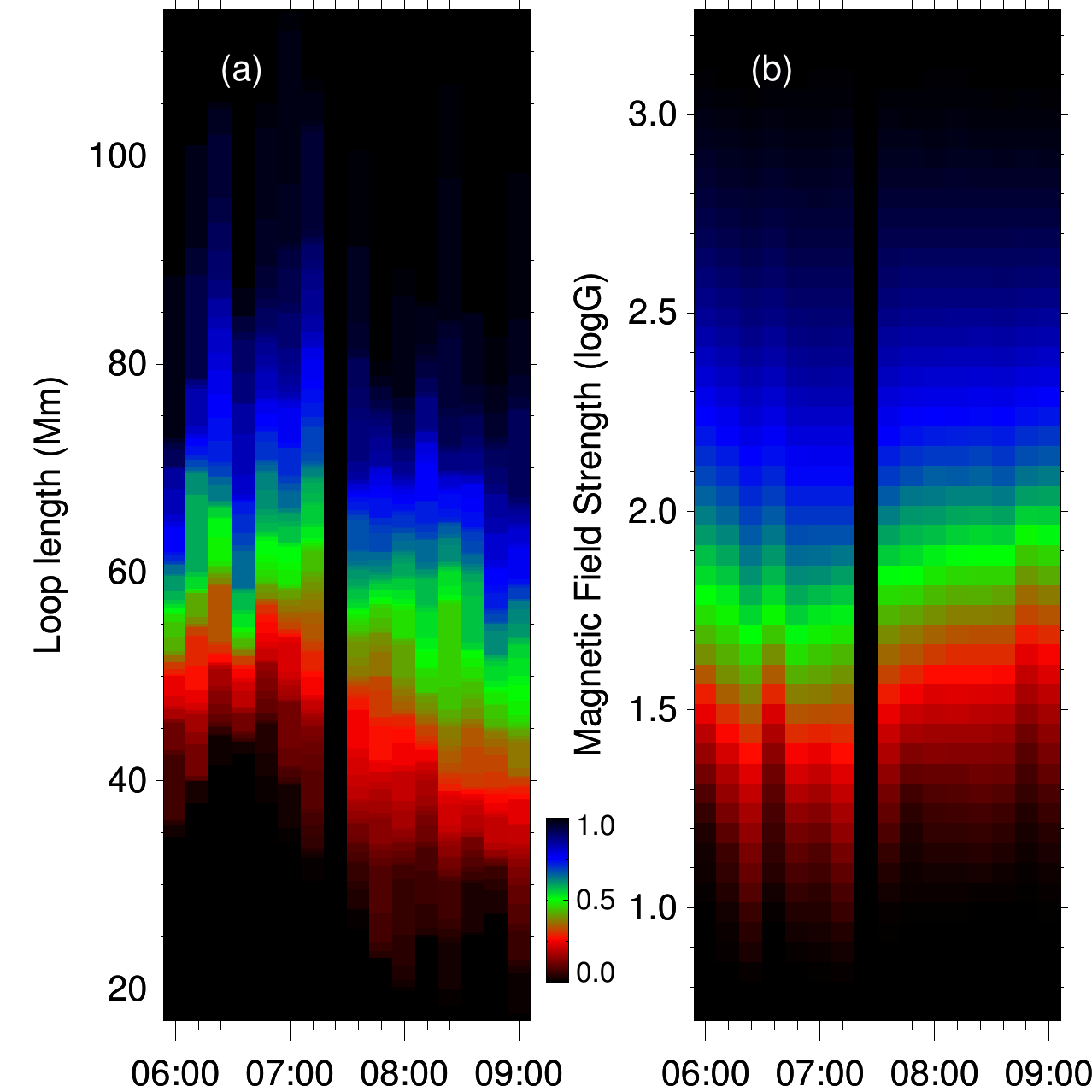}
\caption{\label{lengthevo}Evolutions of the lengths and magnetic field strengths of the extrapolated arcade field lines from 06:00 UT to 09:00 UT. (a) Color coded timeslices of normalised cumulative histograms of the lengths of the arcade field lines. The black gap at 07:24 UT is when Flare II happens, whose extrapolation data is not used. The timeslices at 06:24 UT and 06:36 UT are less reliable (see the text in Section~\ref{contract} for the explanation). (b) Color coded timeslices of normalised cumulative histograms of the magnetic field strengths of all pixels of the extrapolated arcade field lines. A color version of this figure is available in the online journal.} \end{centering}
\end{figure}

The evolution of lengths (and strengths) of the model field lines in the reconstructed arcade from 06:00 UT to 09:00 UT are further explored. We use the same ``timeslices'' technique as in the time-distance diagrams in Figure~\ref{fevolve}, but here in Figure~\ref{lengthevo} each timeslice represents a colour-coded normalised \emph{cumulative} histogram. The black gap at 07:24 UT is when Flare II and arcade contraction happen, thus the extrapolation data is not used. The idea of this figure is to show how the distribution of lengths (and strengths) evolves in time. The black regions at the top and bottom mean that there are no field lines of those lengths there, and the field lines exist in those blue, green and red regions. As we can see, before the flare most of the field lines have lengths between  $\sim$ 40-80 Mm, and after the flare this range shifts down to $\sim$ 30-70 Mm. In addition, before the flare the general trend of the field line lengths is increasing (though a
 relatively strong activity at $\sim$ 06:30 UT, compared to slow evolution in the rest time from 06:00 UT to 07:00 UT, may affect the reliability of the extrapolations at 06:24 UT and 06:36 UT), whereas after the flare it turns to decreasing. The evolution of the field strength of the model arcade in Figure~\ref{lengthevo}(b) shows an opposite trend.

\subsection{Flux Rope and Connectivity Changes}\label{reconnection}
As the overlying arcade and the filament western part share the same expansion and contraction speeds, which can be seen in Figure~\ref{fevolve}(b), the overlying arcade dynamics may be controlled by the filament underneath, or more correctly by its magnetic flux rope. The filament also seems to be the driver of the subsequent flare evolution. Thus it is important to study the change of the filament. As the extrapolation only applies to quasi-equilibrium evolution, we then infer its behaviour from the initial and final extrapolated states.

\begin{figure*}
\includegraphics[width=0.47\textwidth]{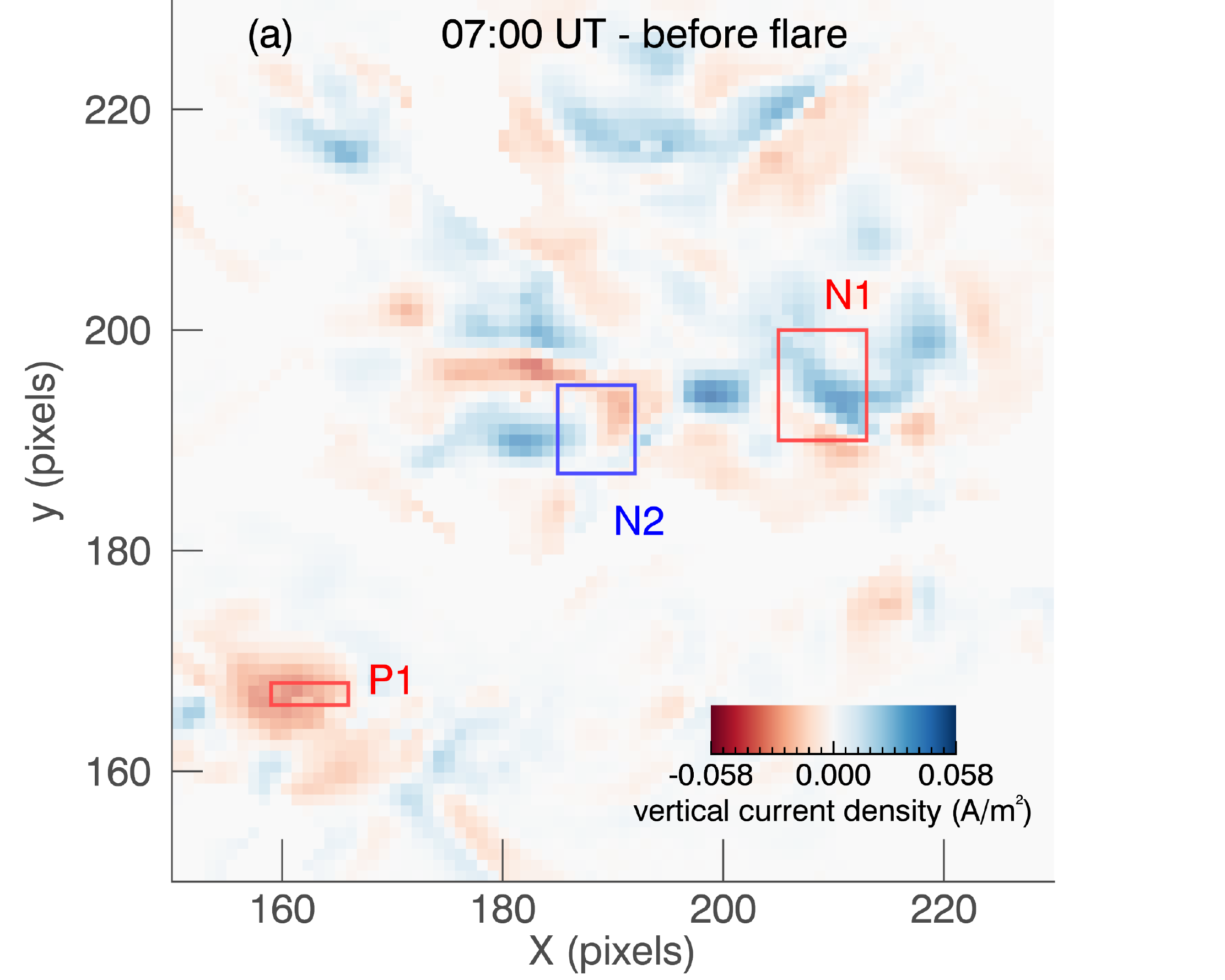}
\includegraphics[width=0.47\textwidth]{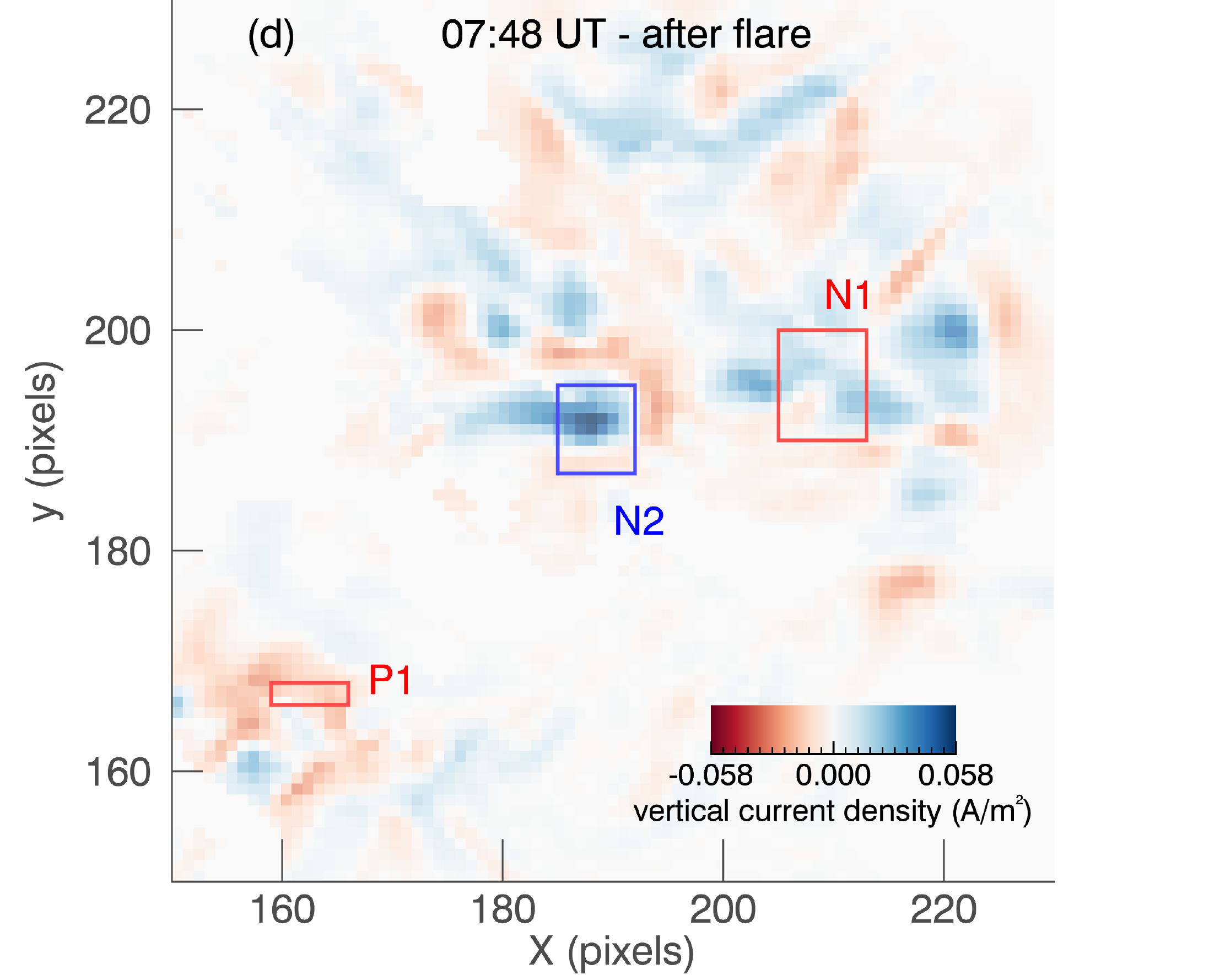}
\includegraphics[width=0.47\textwidth]{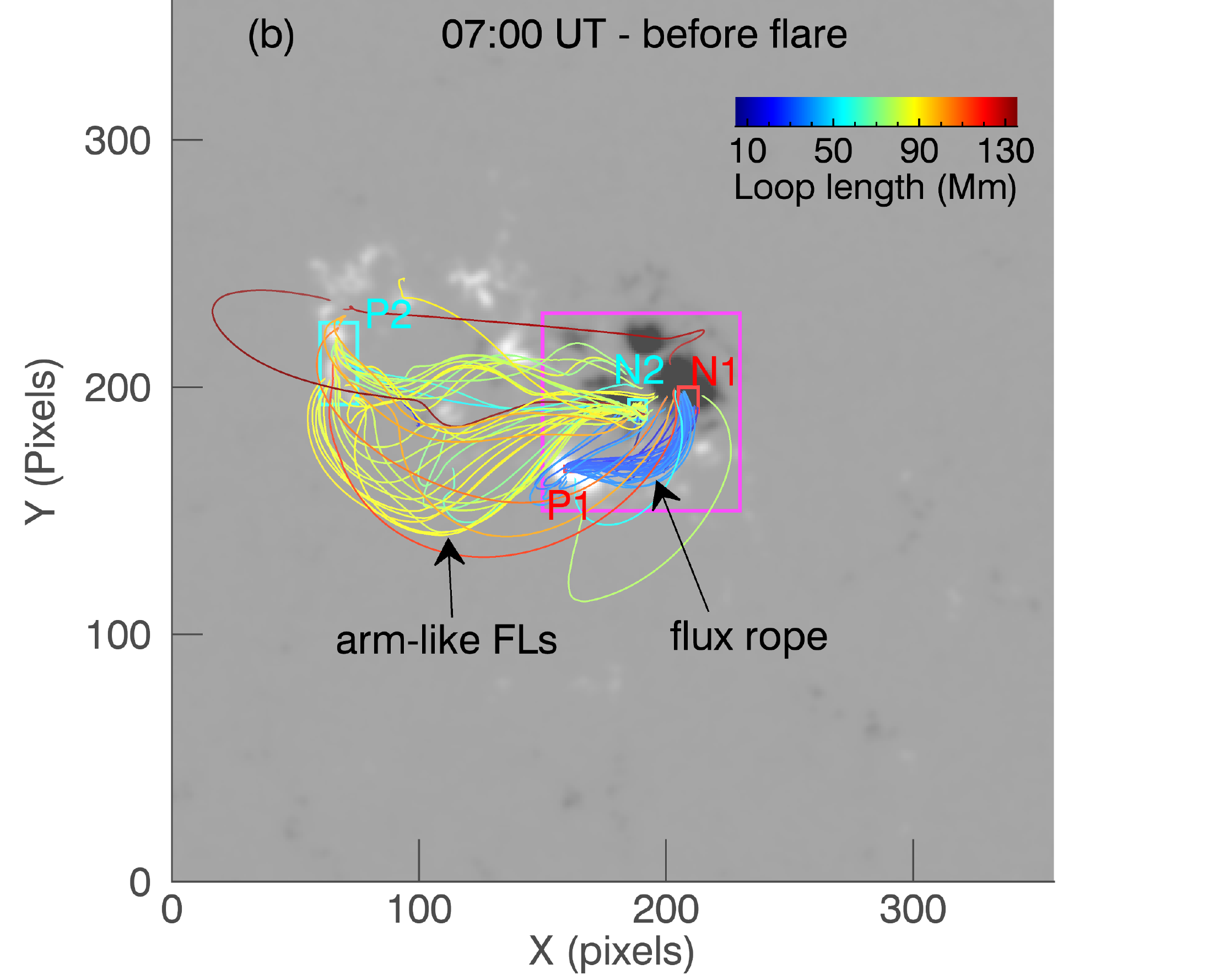}
\includegraphics[width=0.47\textwidth]{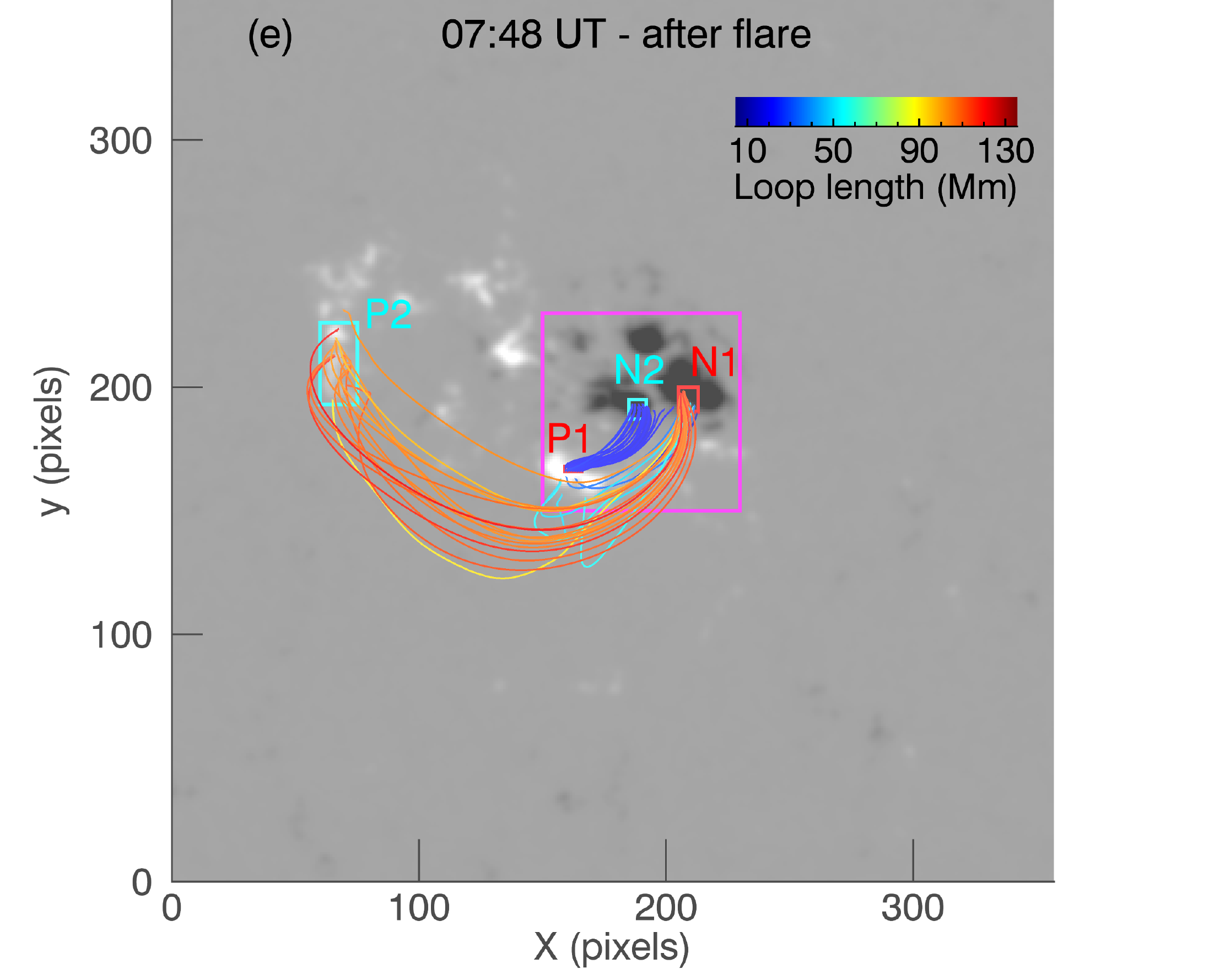}
\includegraphics[width=0.47\textwidth]{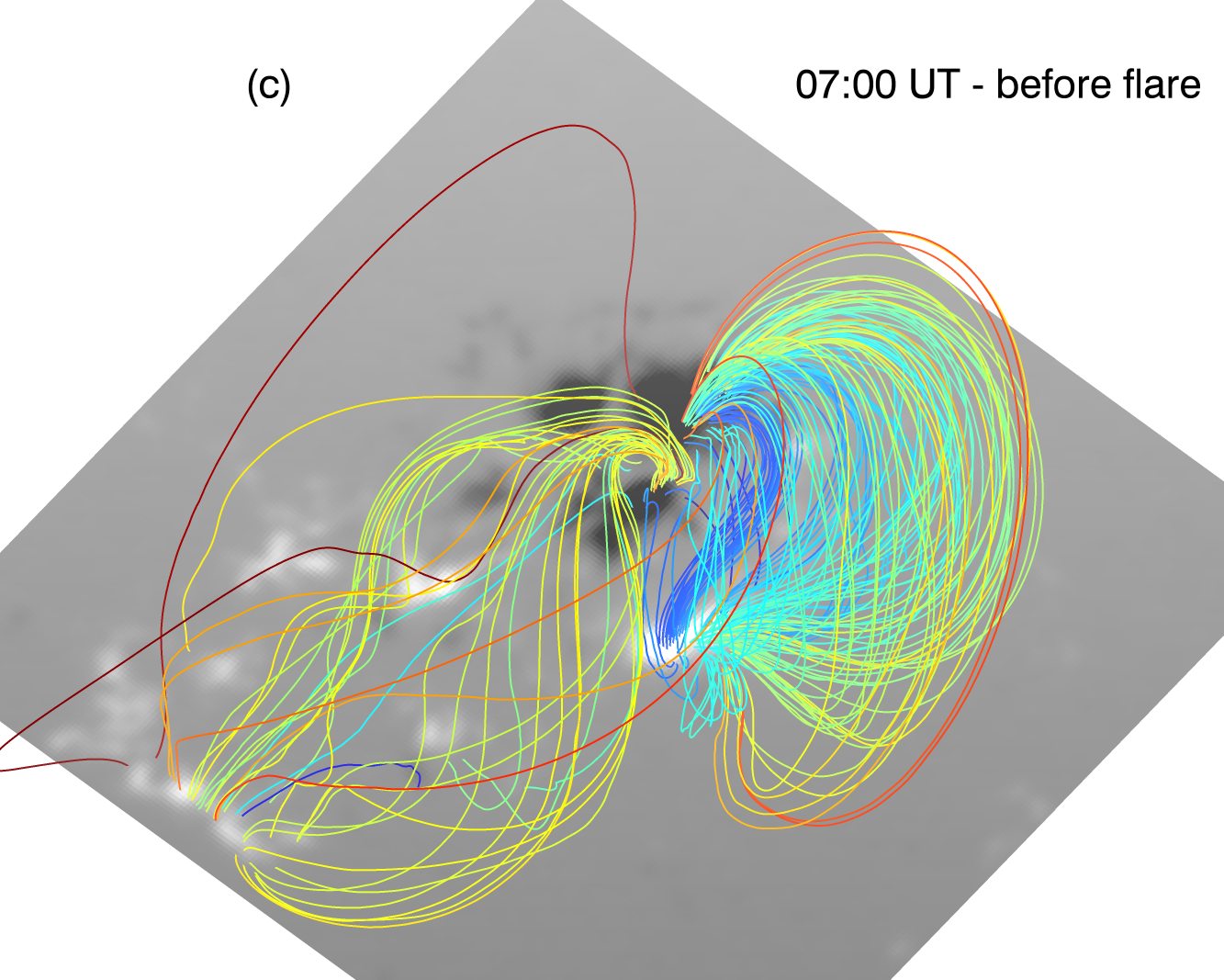}
\hspace{1cm}\includegraphics[width=0.47\textwidth]{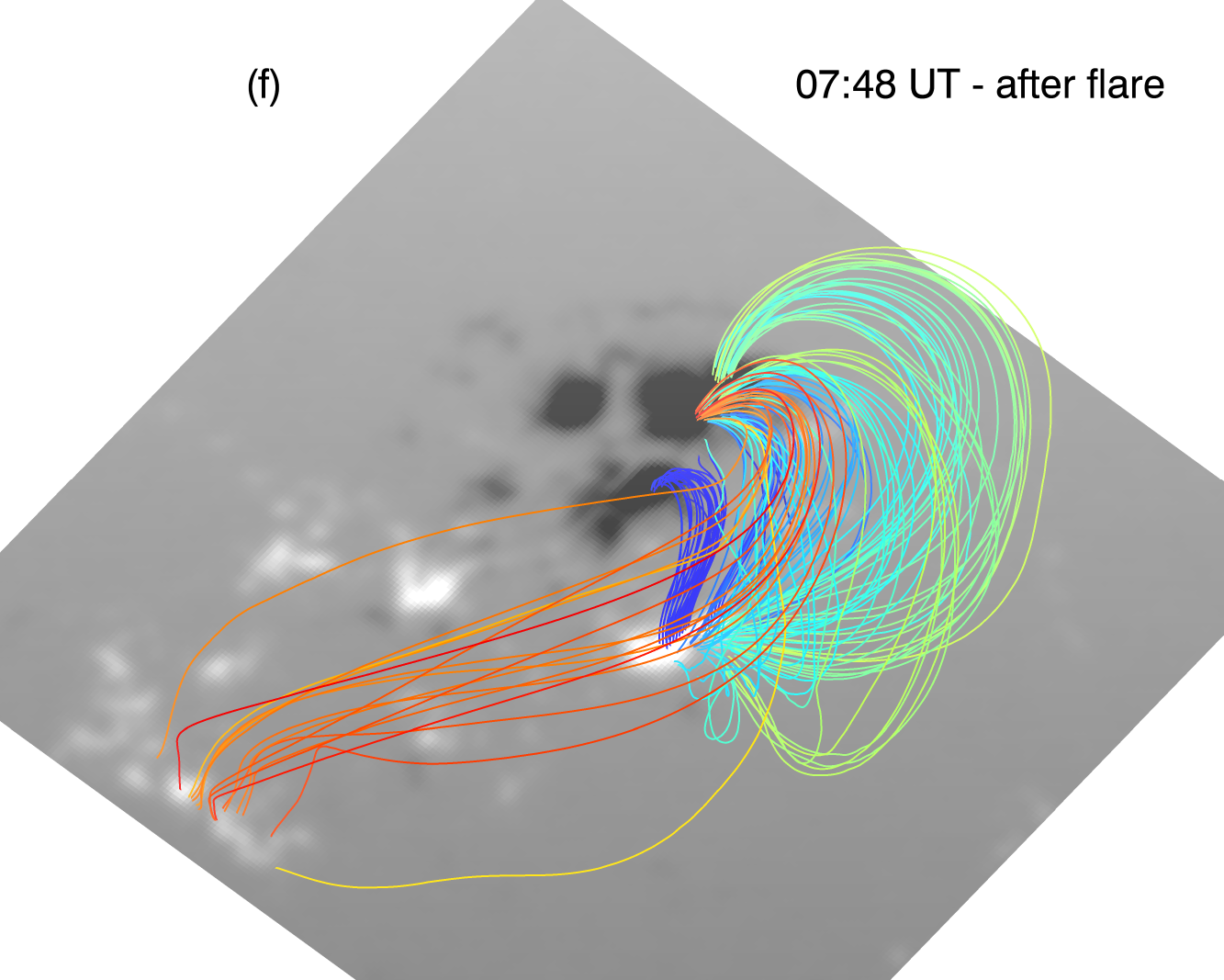}
\caption{\label{rope}Possible flux rope reconnection scenario. (a) Photospheric vertical current density diagram in the magenta square region in (b) at 07:00 UT (before flare). (b) Connectivities at 07:00 UT (before flare). The FOV is approximately the same as in Figure~\ref{structure}. 1 pixel $\approx$ 0.5 arcsecs. (c) 3D view of the connectivities in (b). (d)-(f) Same as (a)-(c), but at 07:48 UT (after flare). The overlying arcade is added in (c) and (f) appearing on the right to show  its relative position and the accompanying implosion. A color version of this figure is available in the online journal.} 
\end{figure*}

At 07:00 UT, before the flare, we find the possible flux rope involved in the activity (blue clustered field lines in Figure~\ref{rope}(b)) in an area P1 of positive polarity and strong vertical currents, seen in Figure~\ref{rope}(a). The rope is very sheared and connected to an area N1 which is just south of the overlying arcade footpoints in the northern negative polarity region. In orientation and size it is very similar to, and could be, the filament seen in  AIA 304 {\AA} (Figure~\ref{filament}(a)). At 07:48 UT, after the flare, we use the same flux rope footpoints in Figure~\ref{rope}(b) at 07:00 UT as the leading footpoints to calculate the new field line connectivities. Figure~\ref{rope}(e) shows that the field lines from P1 are now connected to a closer negative polarity area N2 while those from N1 now connect to the far eastern positive polarity region P2. These two new magnetic systems both become less sheared compared to the original flux rope in Figure~\ref{rope}
(b). The vertical current densities in P1 and N1 meanwhile decrease whereas that in N2 increases. 

To further investigate the change in connectivity, we use the footpoints obtained above in P2 and N2 as leading footpoints, and calculate their connection states before the flare at 07:00 UT. The result in Figure~\ref{rope}(b) - not including the blue clustered flux rope field lines - shows that P2 and N2 are mostly connected by the yellow field lines before the flare, whose profile in the south is very similar to the shape of the expanding arm-like structure seen in AIA 94 {\AA} in Figure~\ref{armstruc}. Hereafter we call these yellow field lines arm-like field lines. As exhibited in Figure~\ref{c4aia}(d) and Figure~\ref{armstruc}, the arm-like structure in 94 {\AA} and the erupting filament in 304 {\AA} accompany each other during the eruption, and they both disappear off the edge of Figure~\ref{armstruc}(i). Thus it may be possible that they reconnect and exchange footpoints during the eruption, leading to a change in the field configuration from that in Figure~\ref{rope}(b) to that in Figure~\ref{rope}(e).

\begin{figure*}
\includegraphics[width=0.47\textwidth]{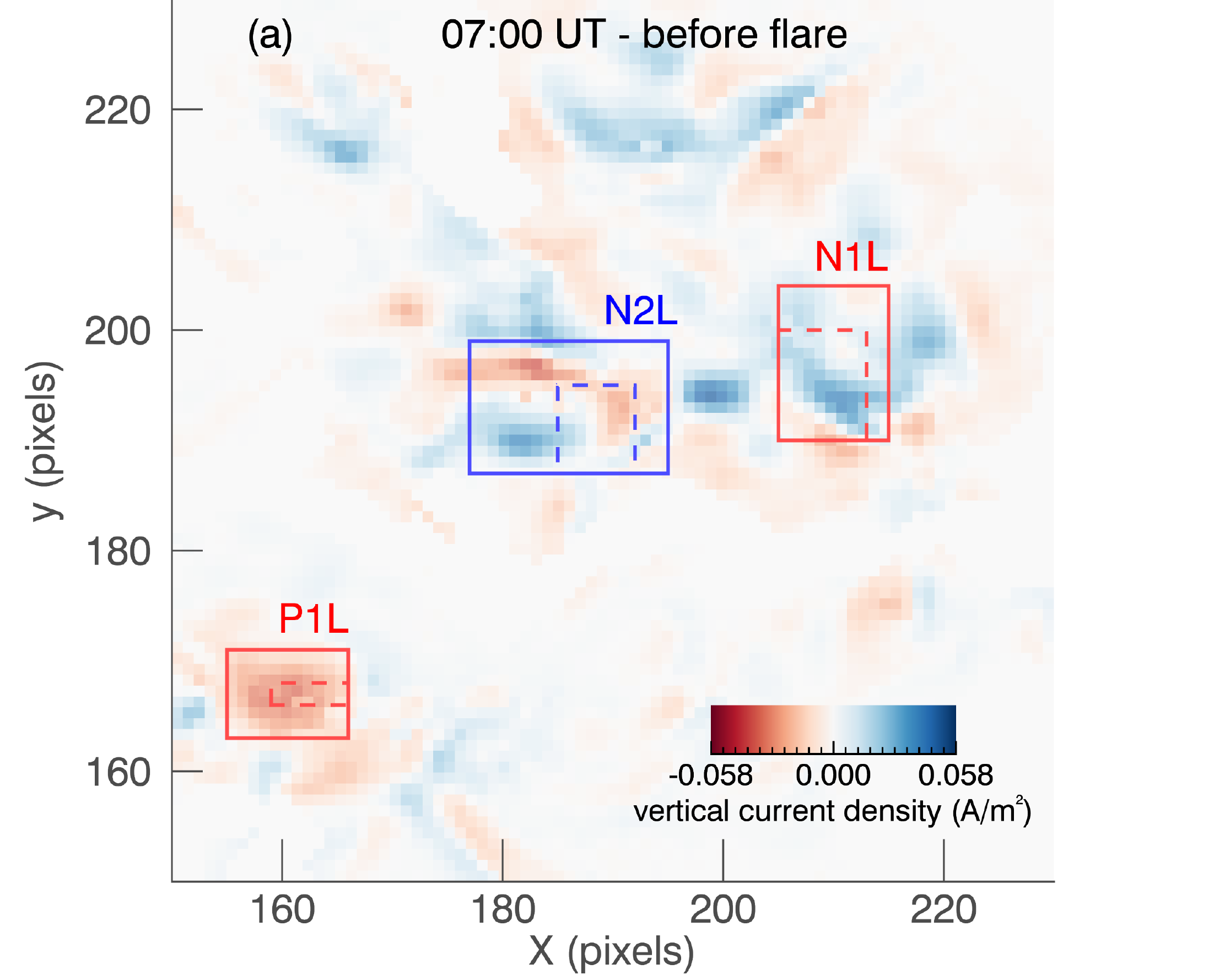}
\includegraphics[width=0.47\textwidth]{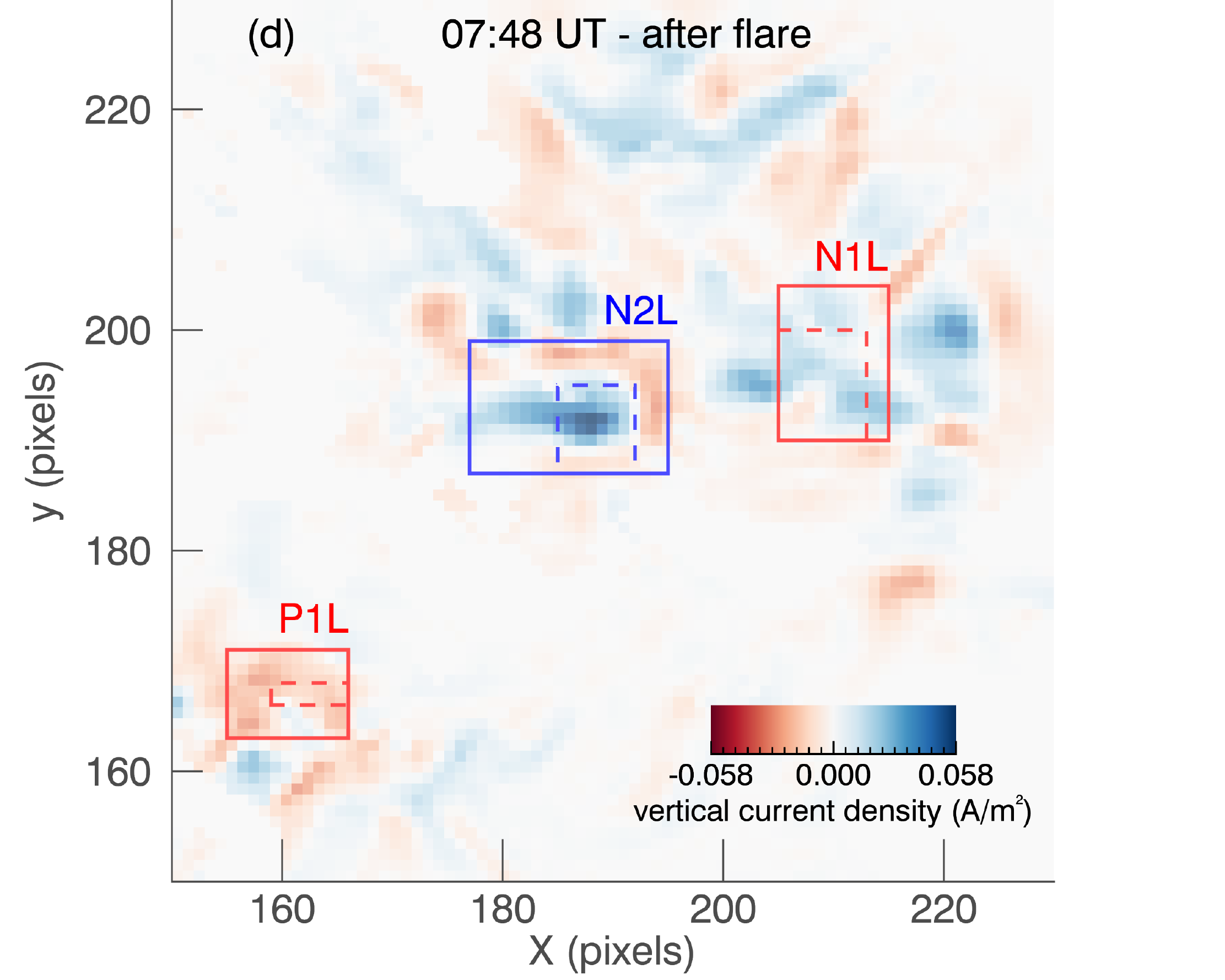}
\includegraphics[width=0.47\textwidth]{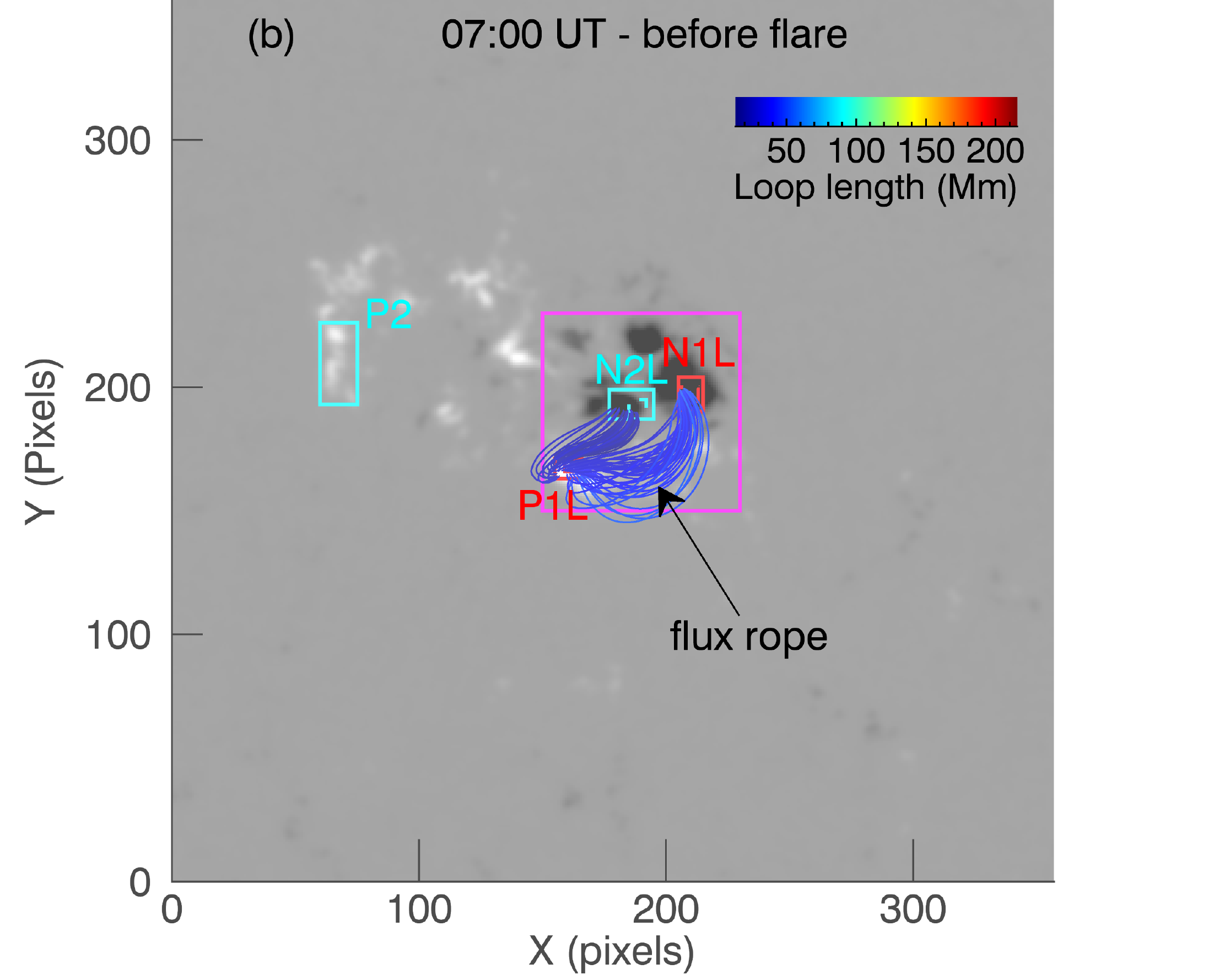}
\includegraphics[width=0.47\textwidth]{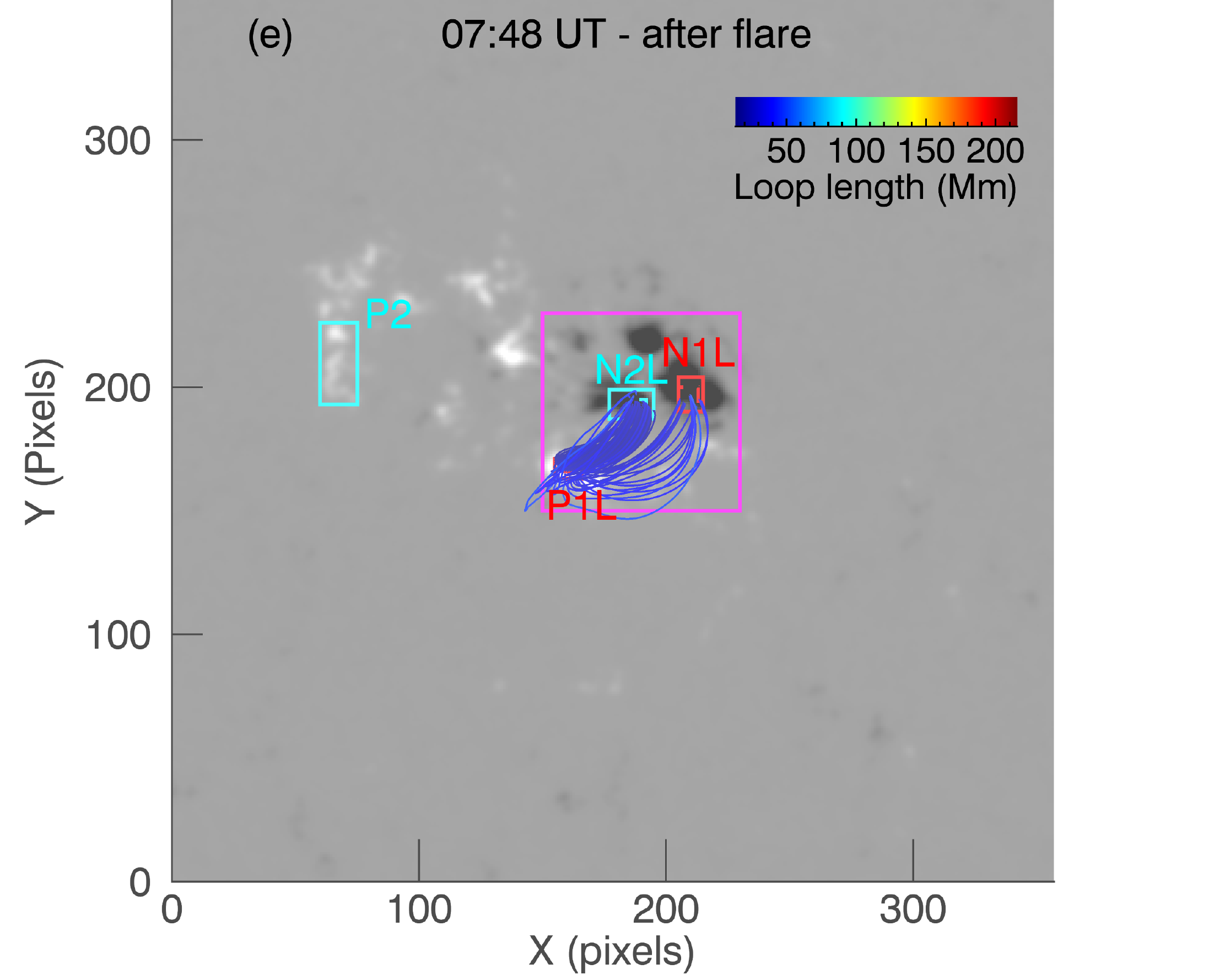}
\includegraphics[width=0.47\textwidth]{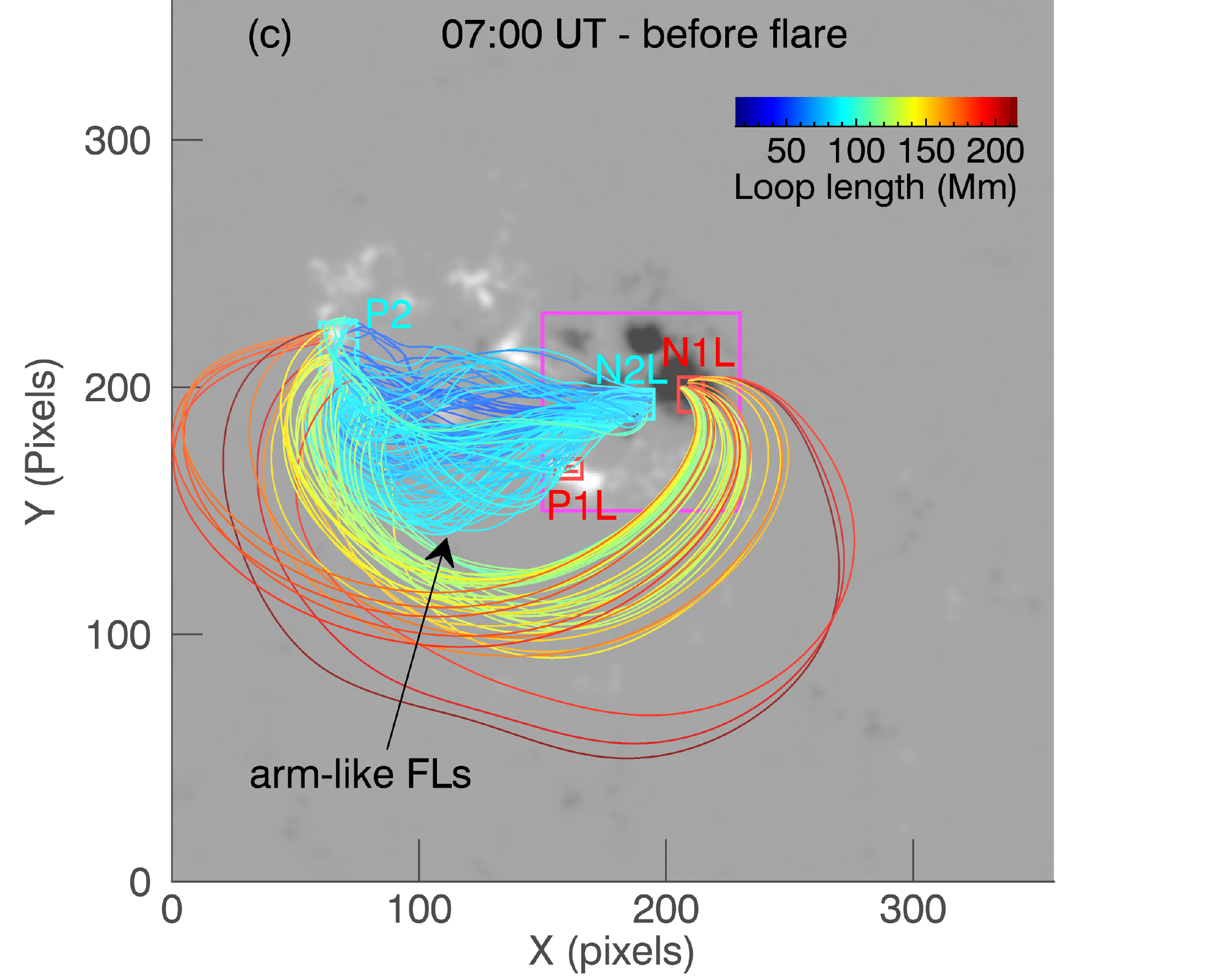}\llap{\includegraphics[height=2.5cm]{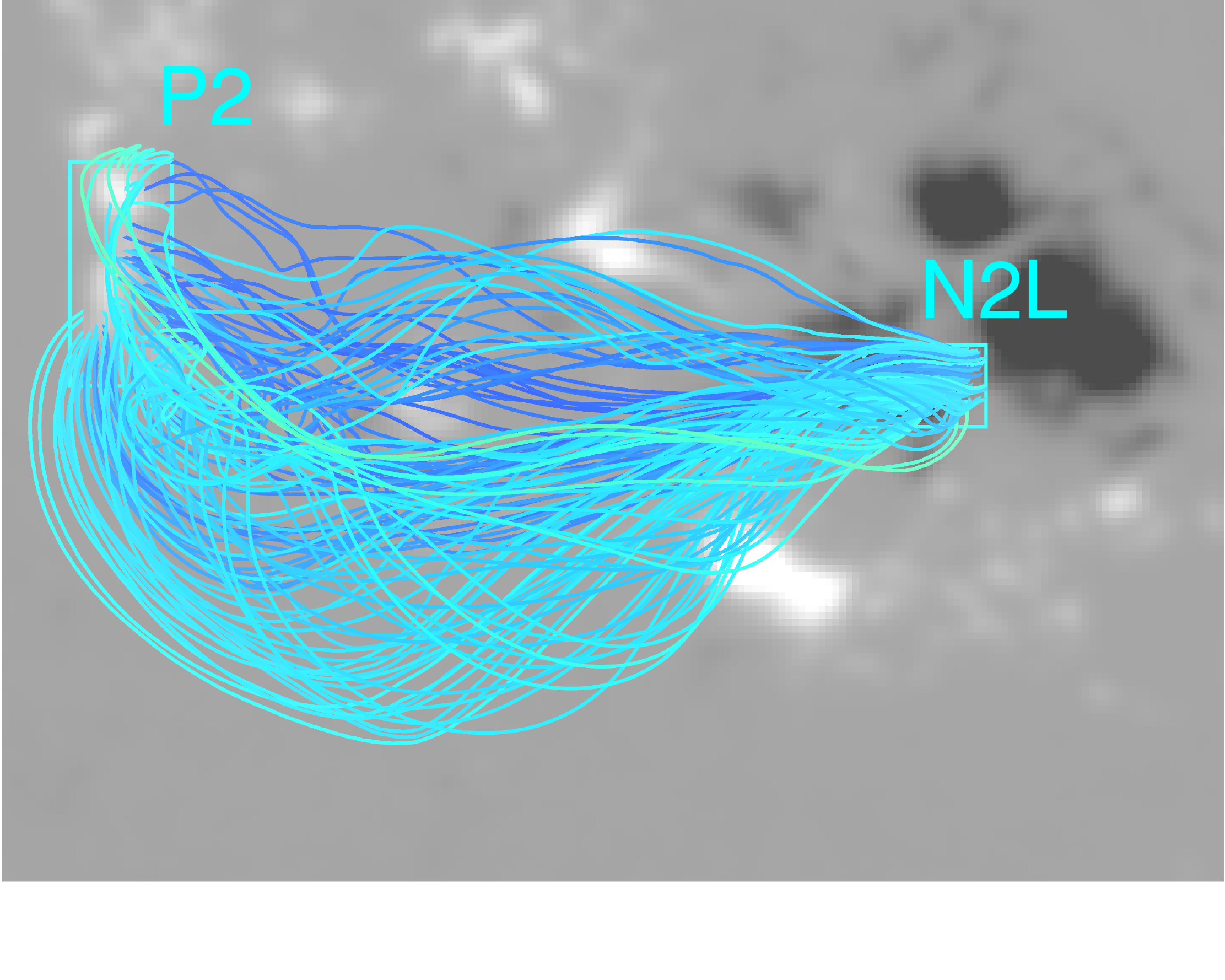}}
\hspace{1cm}\includegraphics[width=0.47\textwidth]{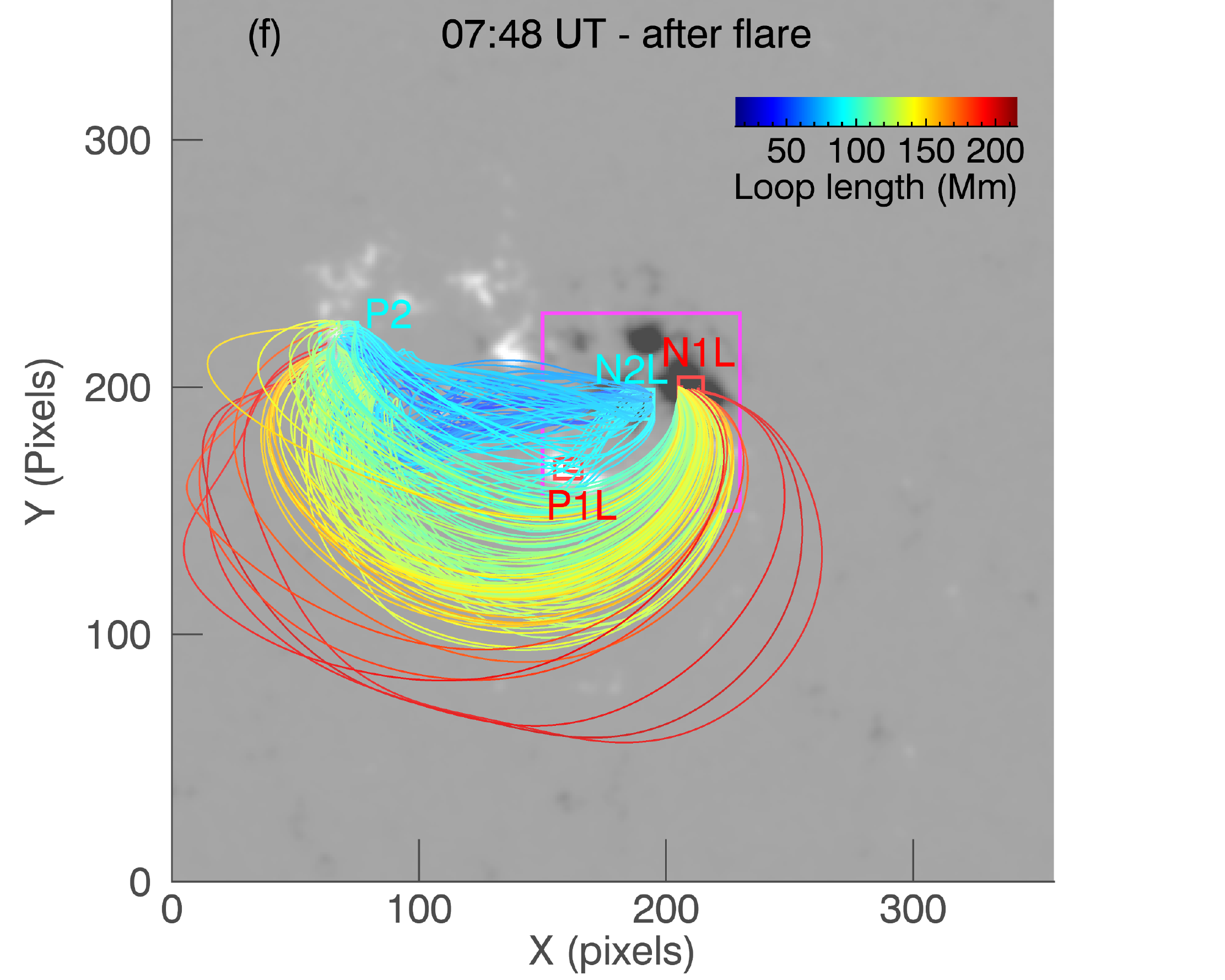}\llap{\includegraphics[height=2.5cm]{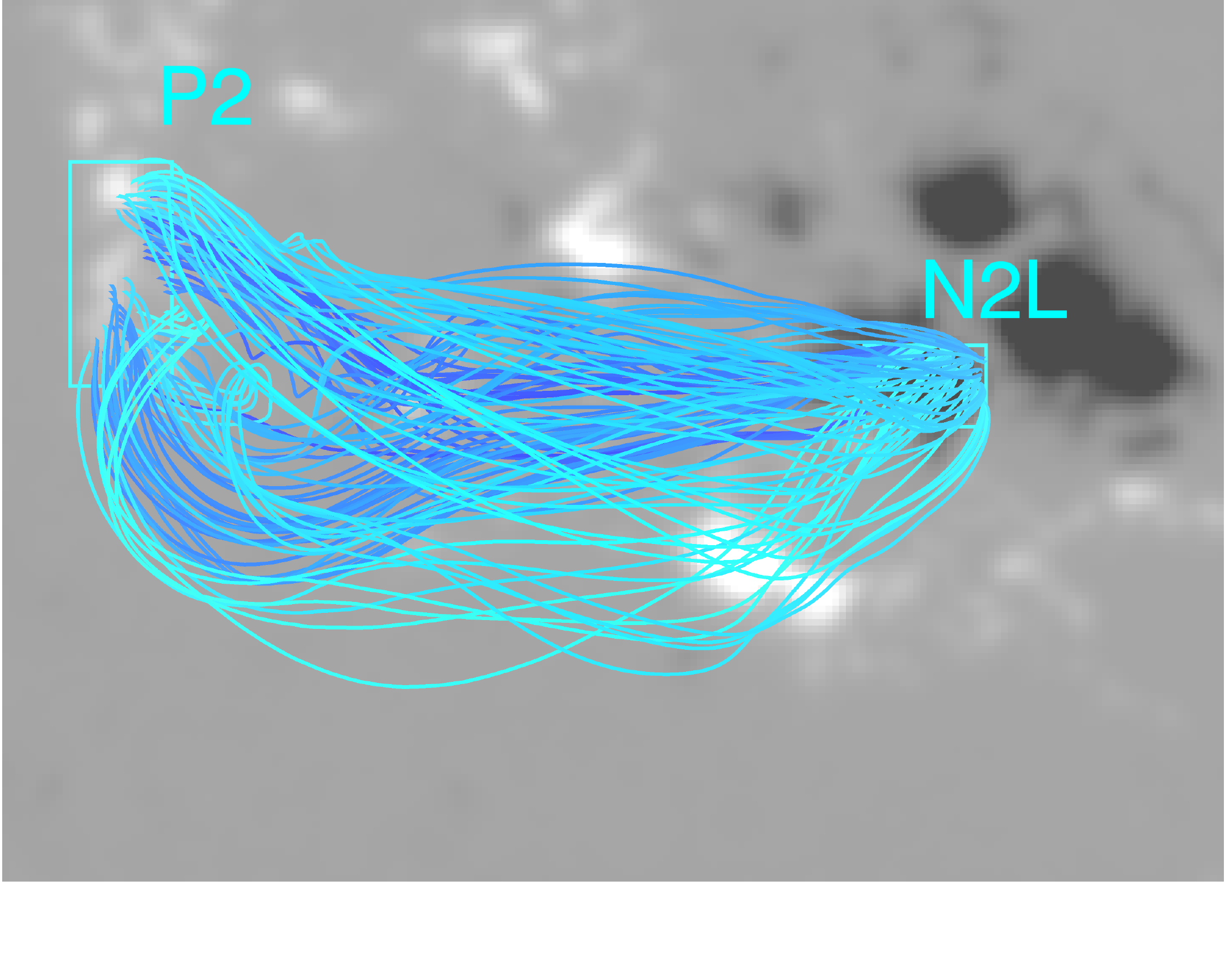}}
\caption{\label{rope2}Connectivity states between the four regions, P1L, N1L, P2, N2L, before and after the flares. (a) Photospheric vertical current density diagram in the magenta square region in (b) at 07:00 UT (before flare). The solid larger boxes are chosen to reduce the influence of possible photospheric magnetic field evolution. The dashed smaller boxes are the original ones in Figure~\ref{rope}. (b) Connectivities from P1L to N1L and N2L at 07:00 UT (before flare). (c) Connectivities from P2 to N1L and N2L at 07:00 UT (before flare). The arm-like field lines are cyan now because the color table scale is changed. (d)-(f) Same as (a)-(c), but at 07:48 UT (after flare). As the connectivities from P2 to N2L in (f) are obscured, we plot them in the inset at the corner of (f), same as in (c). A color version of this figure is available in the online journal.} 
\end{figure*}

In the above analysis, we have only used some specific footpoints in areas P1, N1, P2 and N2 for field line calculation. However, as stated in Section~\ref{contract}, the footpoint identity may change due to photospheric field evolution. Thus in order to make the result more robust, we choose larger areas P1L, N1L and N2L (the solid rectangular regions in Figure~\ref{rope2}(a) and (d)  which are chosen to accommodate similar structures in the photospheric vertical current density diagrams at 07:00 UT and 07:48 UT). We then study the connections between these three regions and P2, calculating all the field lines from P1L to N1L, P1L to N2L, P2 to N1L, and P2 to N2L. Comparing Figure~\ref{rope2}(b) with Figure~\ref{rope2}(e) shows that after the flare, the number of connections\footnote{The number of field lines is generally believed to be a non-physical quantity in a continuous magnetic field. However, as here the measured magnetogram is discrete and only one field line is plo
tted in one pixel of an area $\approx0.5\times0.5$~arcsec, the number of field lines in this situation in fact reflects the bottom boundary area that contributes to the connection between the two regions.} between P1L and N1L decreases, but increases between P1L and N2L. Most of the disappearing connections are the flux rope field lines. In Figure~\ref{rope2}(c) and (d), a similar situation happens with the area P2. The arm-like field lines from P2 to N2L disappear after the flare while the connectivities between P2 and N1L are considerably enhanced. These connectivity changes could be realised by the above proposed possible reconnection between the flux rope and the arm-like field lines. We quantify these changes using the method in \citet{wie2013} to calculate the connected magnetic flux between these four regions at both times. In this method, the flux linking two sources is calculated as the mean of the values obtained taking each source in turn as the leading footpoint r
 egion, with the error given by the half the difference of these values. Before the flare at 07:00 UT the magnetic flux between P1L and N1L is $585.3\pm26.3$ GWb, while after the flare at 07:48 UT it reduces to $192.8\pm58.0$ GWb. The flux between P2 and N2L also declines, from $336.5\pm16.9$ GWb before the flare to $302.2\pm13.7$ GWb after the flare (the reason for this small decrease $\sim10$\% might be that the arm-like field may only account for a small part of the entire connectivities between P2 and N2L, as can be seen by comparing the two insets in Figure~\ref{rope2}(c) and (f), which could result in a relatively small percentage of the total magnetic flux between the two regions). However, the flux between P1L and N2L, and between P2 and N1L, are both enhanced after the flares, from $301.0\pm46.2$ GWb to $787.7\pm97.5$ GWb, and from $251.6\pm55.7$ GWb to $462.8\pm66.2$ GWb, respectively. These flux changes reflect that the connectivity between P1L and N1L and that bet
 ween P2 and N2L are both reduced after the flare, whereas the connectivity between P1L and N2L and the one between P2 and N1L both increase. This could be resulted from the above proposed reconnection between the flux rope and the arm-like structure. This more robust argument increases the likelihood of this scenario.

\section{Discussion} \label{discussion}

\subsection{Evidence for the Implosion}\label{evidence}
\begin{figure}[]
\includegraphics[width=0.495\textwidth]{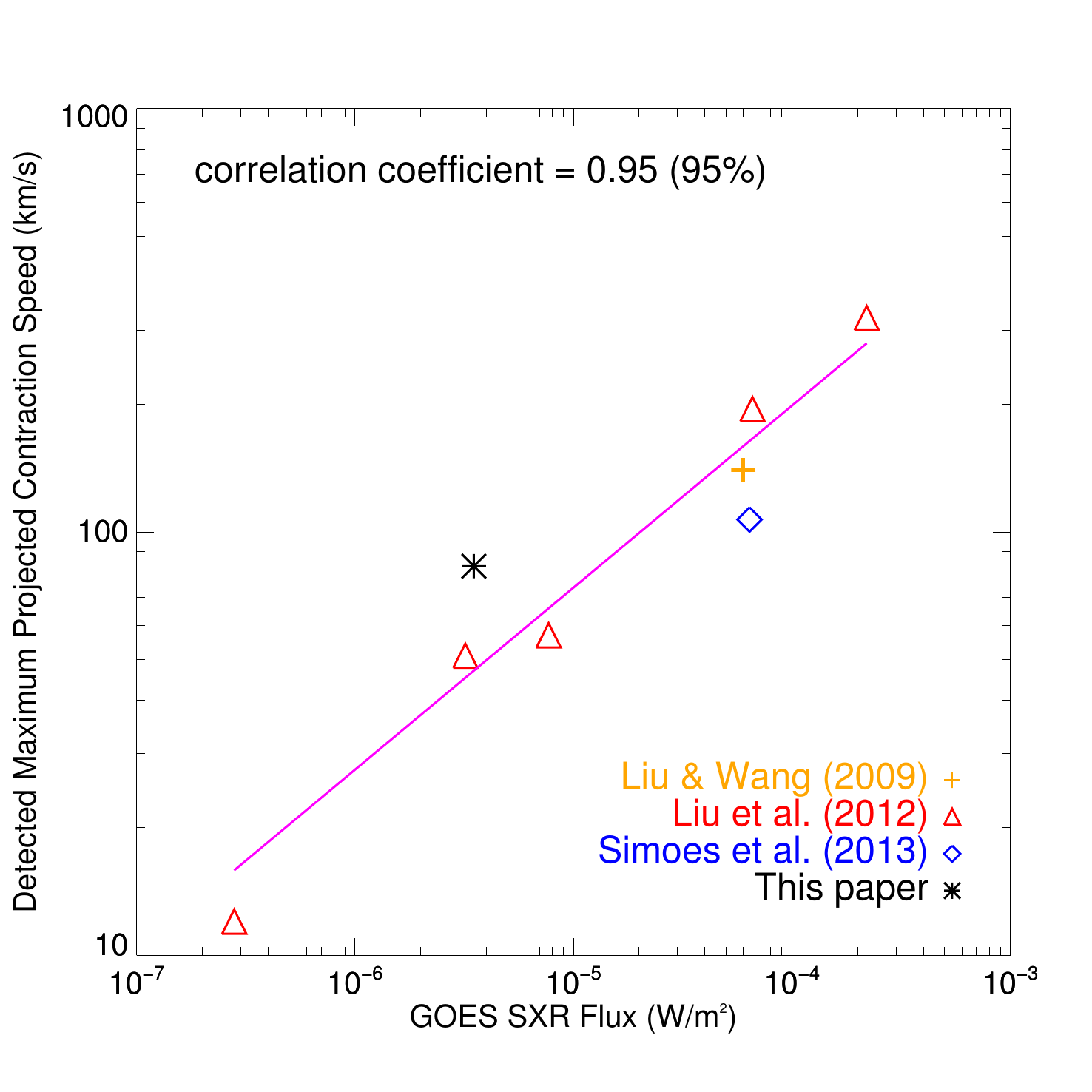}
\caption{\label{fluxspeed}Correlation between the detected maximum projected contraction speed and the SXR flux for 8 disk AR flares, an updated version of \citet{liu2012}.  The magenta line represents the linear regression. The correlation coefficient is 0.95 with a 95\% confidence level. A color version of this figure is available in the online journal.}
\end{figure}

The observed overlying arcade motion shown between ``B'' and ``C'' in Figure~\ref{fevolve}(b) is a contraction without obvious oscillations\footnote{The feeling of oscillations in the 171 {\AA} animation in Figure~\ref{arcfila} might be caused by the gradual brightening of outer contracting loops, which may generate an illusion of the loops bouncing back.}, consistent with a theoretical implosion evolution in which the reduction of magnetic pressure underneath the arcade is slow compared to the arcade loop oscillation period (see Figure 4(b) of \citet{rus2015}). The evidence that this apparent contraction is a real implosion comes from three aspects.
\begin{itemize}
 \item[(i)] In Figure~\ref{fevolve}(b),  the cyan dotted line between ``B'' and ``C'' shows that the arcade apparently contracts by about a half of its original projected height during this period (which can also be seen by comparing Figure~\ref{arcfila}(e) with (h)). An apparent contraction could also be due to a change in loop inclination from a face-on state. However, in this event, if the change were caused only by inclination of the arcade towards the solar disk, the arcade plane would need to incline by about $60^{\circ}$ towards the solar disk in order to satisfy the observed contraction. As the event is close to the disk centre (see Figure~\ref{fullview}), this is quite an unlikely situation (unless the arcade loops can submerge into the photosphere). Thus, inclination only could not account for the observed apparent contraction of the overlying arcade.
 \item[(ii)] The two downwards-moving features between ``B'' and ``C'' in Figure~\ref{fevolve}(b) are nearly parallel to each other during $\sim4$ mins. The simplest explanation is that during this period, as the overlying arcade moves as a whole, its individual loops mostly contract with similar speeds and no dramatic change in inclination (too much change in inclination would cause the two downwards-moving features to converge or diverge). The movement of the arcade in the 171 {\AA} animation in Figure~\ref{arcfila} after 07:22 UT (``B'' in Figure~\ref{fevolve}(b)), which appears to be a moderate inclination superimposed on a major contraction, supports this explanation. Consecutive brightening of the arcade loops at constant projected distance within 4 mins could also give the appearance of the two parallel downwards-moving features, but this would be an unlikely coincidence.
 \item[(iii)] Coronal magnetic field extrapolation provides us with further evidence. As illustrated in Figures~\ref{arcade} and \ref{lengthhist}, the lengths of the overlying arcade field lines are globally shifted to shorter values after the flare. The calculated average projected contraction of the higher and longer field lines of the extrapolated arcade is $\sim4.7$ arcsecs, which is in good agreement with the apparent net contraction $\sim4.5$ arcsecs seen in AIA 171 {\AA} (indicated by the blue arrow in Figure~\ref{fevolve}(b)). In addition, Figure~\ref{lengthevo} shows that before the flare the arcade field line lengths are tending to lengthen, whereas after the flare the trend is decreasing and the global arcade field lengths decrease substantially without restoration for a long time. More compact field after flares has also been found in \citet{sun2012} and \citet{tha2016}.
\end{itemize}
Even though reported magnetic field implosions are still rare, implosions could in fact happen frequently. Sometimes it may be their relatively small displacements in small flares, compared to nearly simultaneous violent eruptions or CMEs, that make them hard to recognise. As in our event, if it were not for the first expansion phase that inflates the overlying arcade, the final apparent net contraction $\sim4.5$ arcsecs  would be relatively difficult to discover. However, as the released flare energy increases, implosion could be more noticeable, as in the M6.4 flare where a displacement of $\sim25$ arcsecs has been seen (\citeauthor{sim2013a} \citeyear{sim2013a}) and the X2.2 flare $\sim40$ arcsecs (\citeauthor{gos2012} \citeyear{gos2012}; \citeauthor{liu2012} \citeyear{liu2012}; \citeauthor{sun2012} \citeyear{sun2012}). Moreover, Figure~\ref{fluxspeed} implies that the maximum contraction speed may also correlate with the released energy level.

\begin{figure*}
\includegraphics[width=0.495\textwidth]{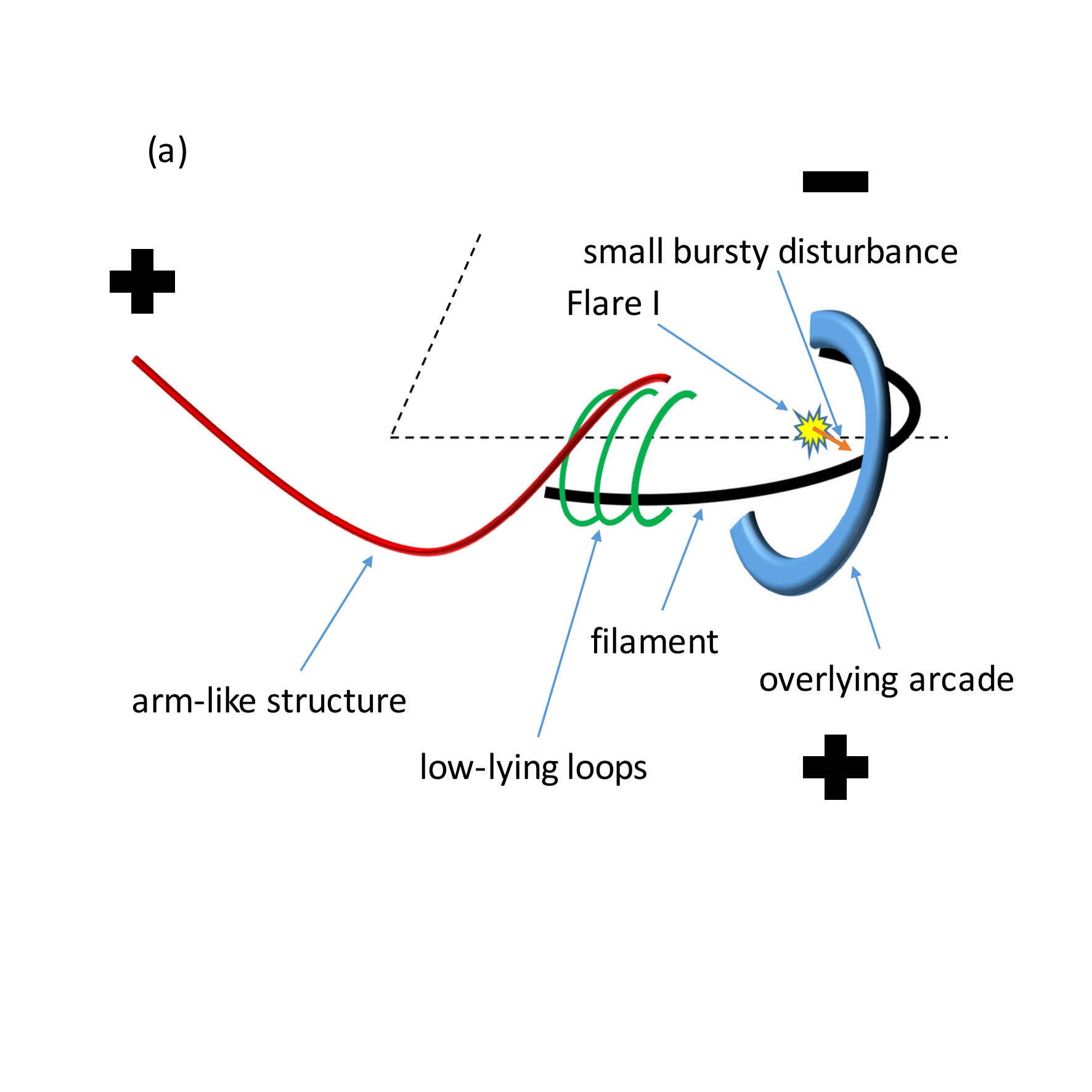}
\includegraphics[width=0.495\textwidth]{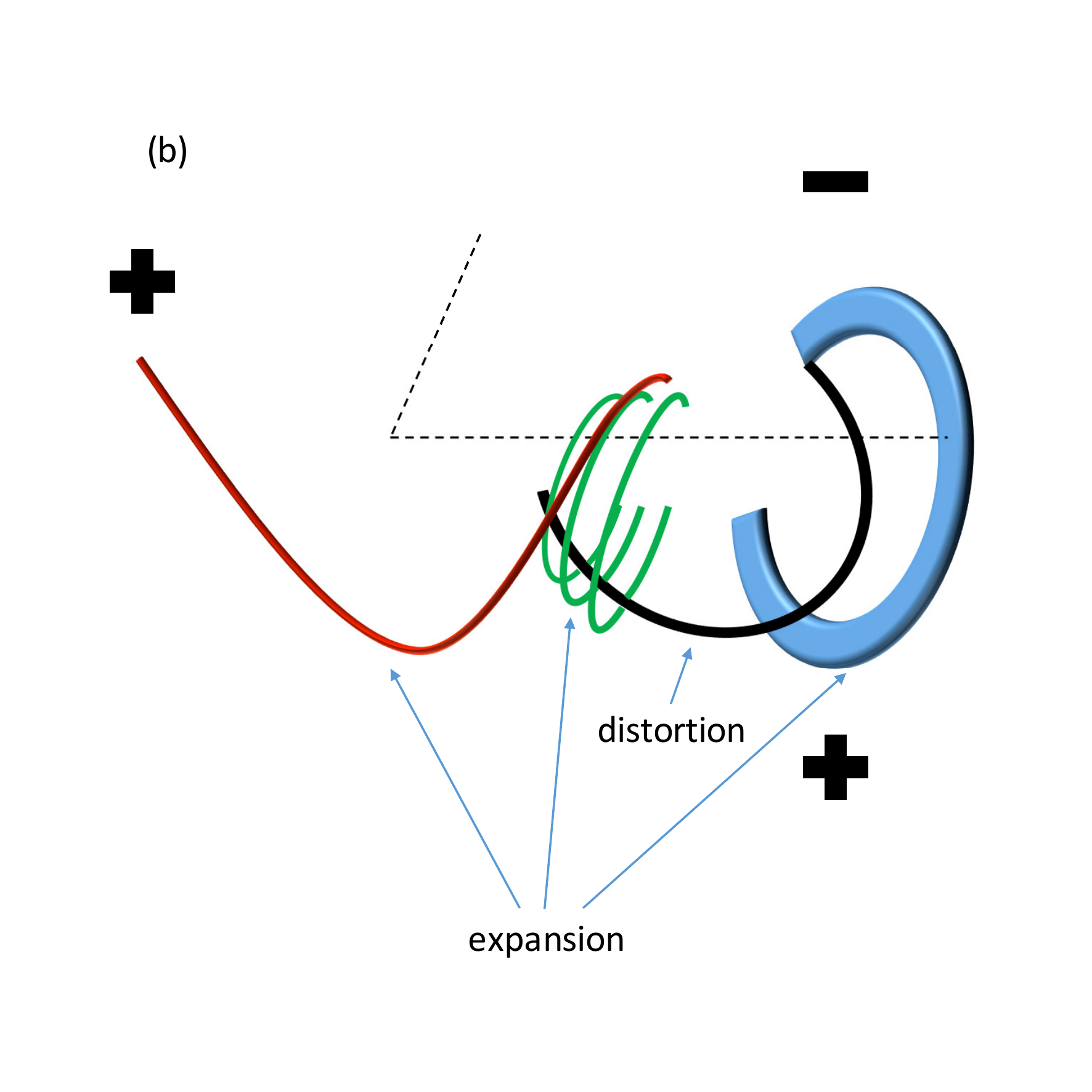}
\includegraphics[width=0.495\textwidth]{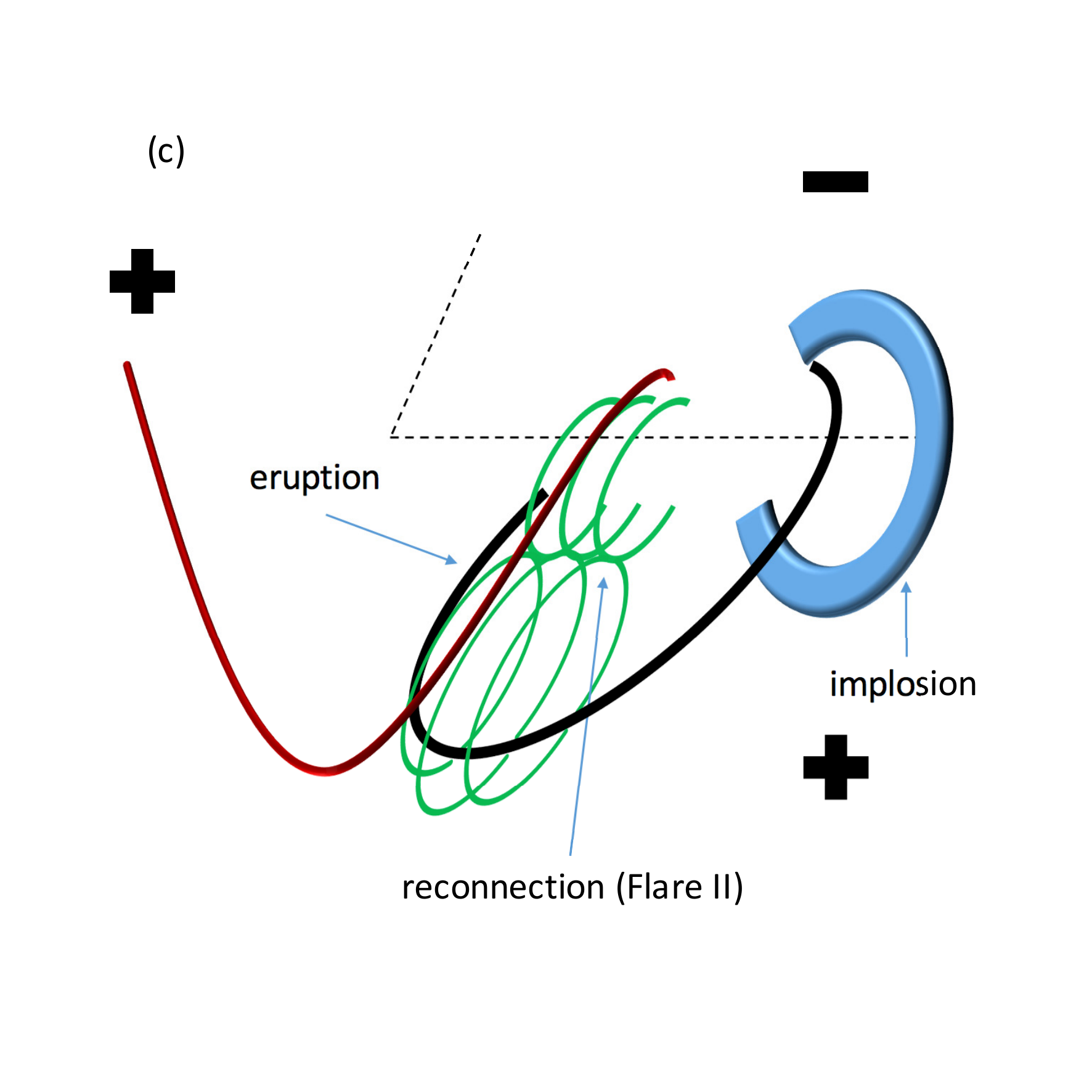}
\includegraphics[width=0.495\textwidth]{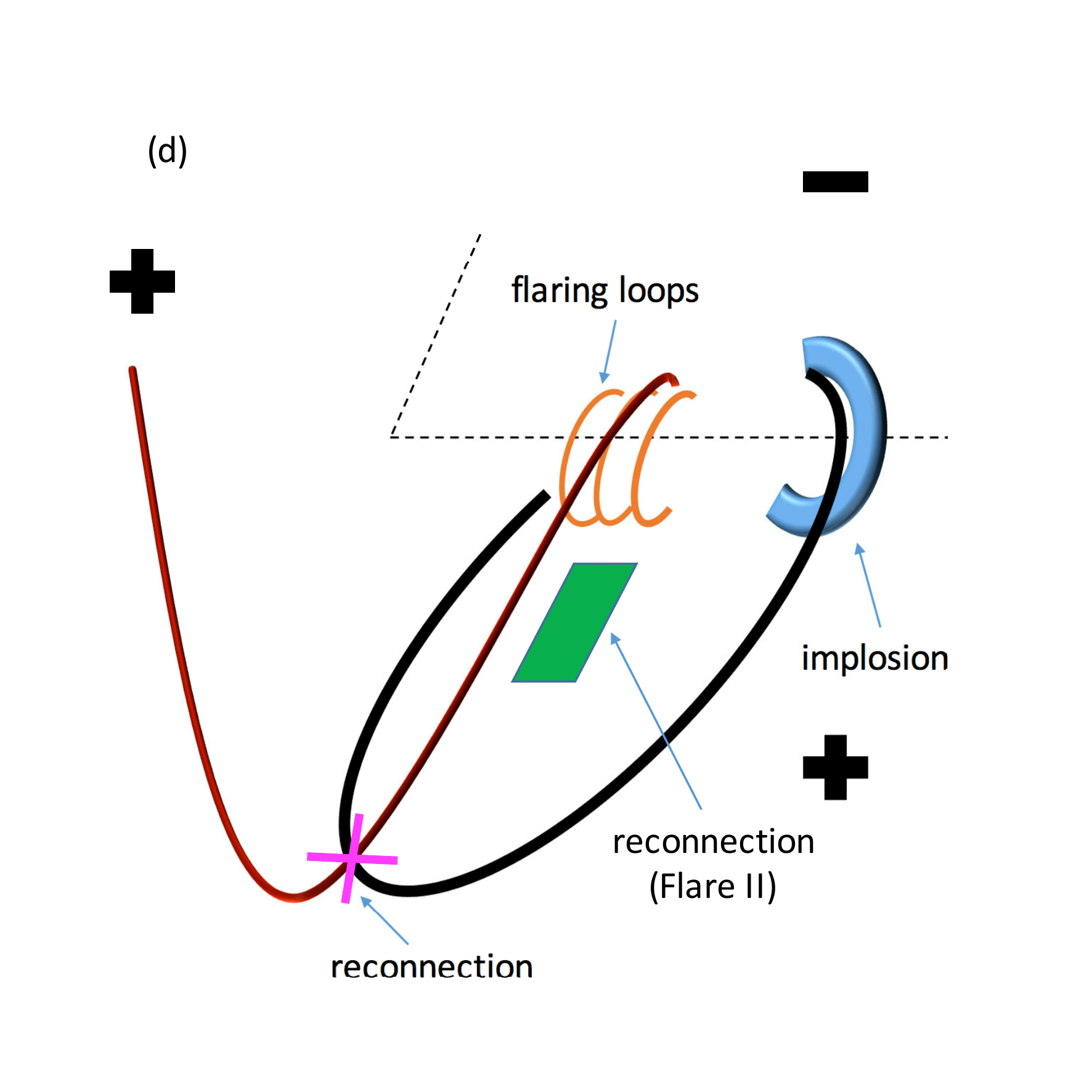}
\caption{\label{cartoon}Cartoons show our understanding of the flare evolution. The large bold ``+'' and ``-'' signs in each image represent positive and negative polarity regions, respectively. (a) Flare I disturbs the filament's western part. (b) Filament distortion phase (with the overlying arcade expansion). (c) Filament eruption phase (with Flare II and the overlying arcade implosion). (d) Further eruption of the filament (with Flare II, the overlying arcade implosion, and the possible reconnection between the filament and the arm-like structure). The green rectangular region in (d) represents that there is still a current sheet reconnection beneath the erupting filament, like in (c), which is used to make the image easier to see.}
\end{figure*}

\subsection{Possible Scenario for the Overall Evolution}\label{scenario}
Table~\ref{timing} shows that the observed evolution consists of four main processes: Flare I, the filament distortion and eruption, Flare II, and the overlying arcade expansion and contraction. As described in Section~\ref{observation}, they exhibit intimate relationships both in time and space. After synthesising the observations and extrapolation results in Section~\ref{observation} and \ref{extrapolation}, in Figure~\ref{cartoon} we illustrate our understanding of the event evolution, mainly in the framework of the metastable eruption model \citep{stu2001}, the implosion conjecture \citep{hud2000}, and the standard ``CSHKP'' model of two-ribbon flares \citep{car1964,stu1966,hir1974,kop1976}. Possibly due to the perturbation produced by Flare I, the initially metastable filament brightens and becomes unstable (Figure~\ref{cartoon}(a)). The overlying arcade restricts the filament from erupting, so it has to distort, with a bump propagating from west to east (Figure~\ref{cartoon}(b)) representing transport of free magnetic energy from an environment with a stronger surrounding field, to a weaker one. When the bump (free energy) propagates through the arcade plane, the arcade expands as a  consequence (Figure~\ref{cartoon}(b)). As the bump propagates further to the east, the filament's eastern part suddenly erupts nonradially, possibly due to an ideal MHD instability (Figure~\ref{cartoon}(c)). This simultaneously causes the overlying arcade to contract according to the implosion conjecture, and Flare II to happen through reconnection (Figure~\ref{cartoon}(c)). As the filament continues to erupt, the arcade contracts further (Figure~\ref{cartoon}(d)). In the following, we will explain the scenario in more detail.

\subsubsection{Scenario for Flare I, the Filament Distortion and Eruption, and Flare II}\label{scefila}
A twisted flux rope anchored below a magnetic arcade can stay in a metastable state, but following a large disturbance, e.g., produced by a nearby flare, could become unstable and rupture through the arcade, leading the system to a lower energy state \citep{stu2001}. At the beginning of our event the magnetic system may be in a metastable state which is then disrupted, possibly by Flare I at the filament's western part (Figure~\ref{cartoon}(a) shown in Figure~\ref{filament}(b) and (c)). The disturbed and brightened western part of the filament is restrained against erupting outwards by the overlying arcade field. The filament instead distorts and a bump or bend in the field, which we  associate with free energy, propagates from west to east (Figure~\ref{cartoon}(b)) where the field is weaker as shown by extrapolations. As the free magnetic energy is transported through the arcade plane,  the arcade is pushed upwards due to the enhanced underlying magnetic pressure. This could
 account for the synchronism of the start of expansion of the filament's eastern and western parts, and the overlying arcade at $\sim$ 07:15:40 UT, revealed by the timeslices in Figure~\ref{fevolve}(a) and (b). As the bump propagates further, close to the filament's eastern footpoint, and sweeps across cut 1 and 2 of Figure~\ref{structure}(b) we expect that the filament's western part would contract while the eastern part expands, corresponding to their observed dynamics in Figure~\ref{fevolve}(a) and (b) between $\sim$ 07:20 UT and 07:22 UT \footnote{Only the filament's western part tracked by cut 2 has a contraction between $\sim$ 07:20 UT and 07:22 UT, while part of the bump still supports the overlying arcade during this time. This can explain the delay of the start of contraction of the overlying arcade at $\sim$ 07:22 UT instead of $\sim$ 07:20 UT, i.e., the asynchronism of the start of contraction of the overlying arcade and the filament western part in Figure~\ref{fevolve}(b).}. The arcade would meanwhile be pushed aside by the filament's western part, and incline more towards the solar disk. At the end of this distortion, the filament's western part also appears compressed (Figure~\ref{arcfila}(d)), possibly caused by the strong downward tension of the overlying arcade field in the west and the weaker confinement of the low-lying loops on the growing filament's bump in the east during the persistent distortion. 

The dramatic acceleration and eruption of the filament's eastern part (Figure~\ref{cartoon}(c), corresponding to the observation at $\sim$ 07:22 UT in Figure~\ref{arcfila}(e)) may be due to the torus instability \citep{kli2006} because of the weaker magnetic field in the expanding eastern low-lying loops, or the kink instability \citep{sak1976,rus1996} due to squeezing of the filament, or both. The surrounding field could then be highly stretched to form a current sheet beneath the erupting filament producing Flare II, as in the standard ``CSHKP'' model of two-ribbon flares.

\subsubsection{Scenario for the Overlying Arcade Contraction}\label{scontract}
When the filament erupts at $\sim$ 07:22 UT, the overlying arcade contraction also starts immediately,  shown in Figure~\ref{arcfila}(e) to (i).  As demonstrated in Section~\ref{evidence}, it is very likely to be a real implosion, due to reduced magnetic energy underneath the arcade. \citet{rus2015} theoretically demonstrate three implosion types, with two having oscillations and the third not. In our event, as shown in Figure~\ref{fevolve}(b), no obvious oscillations have been detected, so it belongs to the ``gradual energy release'' situation (see Figure 4(b) of \citeauthor{rus2015} \citeyear{rus2015}) in which the underlying magnetic energy is released slowly compared to the loop's oscillation period. After carefully inspecting Figure~\ref{arcfila}(d) to (f) and the 171 {\AA} animation in Figure~\ref{arcfila} between $\sim$ 07:21 UT to 07:25 UT, we propose two reasons why the energy release is gradual. Both reflect magnetic energy transfer out of the arcade plane.
 
\begin{itemize}
\item[(i)] The filament erupting outwards from beneath the arcade would enhance the magnetic field to the east of the arcade, which creates a larger magnetic pressure that pushes the arcade to incline towards the solar disk. The relative positions of the filament and the arcade would change, and the interface between the filament's western end and the arcade's southern leg would gradually slip from in the arcade plane to above it (see Figure~\ref{cartoon}(b) to (d)), which means that the component of the magnetic pressure exerted by the filament's western leg in the loop plane would be gradually reduced.
\item[(ii)] As the filament stretches outwards, its magnetic energy is transformed into kinetic and gravitational energy of the erupting plasma \citep{sch2015}. The magnetic energy per unit length would then decrease, manifested by reduced magnetic twist per unit length (see equation 2.2 of \citeauthor{stu2001} \citeyear{stu2001}). This can further reduce the component of the magnetic pressure parallel to the loop plane provided by the filament's western leg.
\end{itemize}

The timescale for these two effects could be such that the overlying loops do not oscillate. The final net contraction seen in both observation (Figure~\ref{fevolve}(b)) and extrapolation (Figure~\ref{arcade} and \ref{lengthhist}) means that finally the field underneath the arcade has a lower magnetic energy density/pressure. 

The rapid contraction of the inner arcade loops occurs only during the rise of Flare II's impulsive phase (between ``B'' and ``C'' in Figure~\ref{fevolve}), as seen in other two events reported by \citet{sim2013a} (see its Figure 4), and by \citet{gos2012} and \citet{sun2012}. This also indicates that the contraction is indeed not directly caused by the flare energy release/conversion, otherwise we would expect a comparable contraction in the declining part of the impulsive phase when the energy dissipated, as the energy content of non-thermal particles producing the HXR flux is comparable. However, the contraction is still related to the flare in that the impulsive phase is associated with the filament eruption out of the AR core.

\subsubsection{Scenario for the Possible Filament Reconnection}
As illustrated in Figure~\ref{c4aia}(d) and Figure~\ref{armstruc}, there is also an arm-like structure accompanying the filament eruption. The extrapolation results in Section~\ref{reconnection} show that they could reconnect with each other and exchange their footpoints during the eruption process, to form a less sheared configuration (compare Figure~\ref{rope}(b) with (e)). The cartoon of Figure~\ref{cartoon}(d) illustrates this possible filament reconnection scenario (the exact reconnection location is uncertain). This could contribute to Flare II to some extent in the late erupting phase, but since the filament in the late erupting phase in 304 {\AA} is too weak to track, it cannot be confirmed by the present observations. However, the reconnection of an erupting filament to a far distant area has been observed in 304 {\AA} in another event by \citet{fil2014} (especially see their movie 3, similar to our event). \cite{li2016} have also recently reported an erupting filament reconnecting with a nea
rby coronal structure.

\section{Conclusions} \label{conclusions}
AIA observations and NLFFF extrapolations point to the well-observed contraction of the overlying arcade during the filament eruption in flare SOL20130619T07:29 being a real implosion rather than an inclination effect. We interpret the implosion as due to magnetic energy transfer out of the arcade plane in the filament eruption process rather than due to local magnetic energy dissipation in the flare. The final net contraction of the arcade reflects the permanent change of magnetic pressure underneath the arcade. This event implies that filament movement or eruption can make overlying field expand or erupt as observed in many events, but also is able to simultaneously implode peripheral or unopened overlying field due to reduced magnetic pressure underneath. This event appears to demonstrate one of the ways in 3D to open the overlying field without violating the Aly-Sturrock hypothesis, that is, ``partial opening of the field'', which allows the field to open in one part of t
he region and to implode in another.

The event is interesting in terms of the diversity of processes involved and their close relationships in space and time. The proposed scenario for its evolution has two main implications: (1) the uneven confinement of a filament by overlying field can force energy transfer through the region, with filament distortion preceding a dramatic and probably asymmetric eruption through a ``weak spot''. To identify such locations, measures of the field confinement such as the decay index \citep[e.g.,][]{liu2008} need to be examined from point to point in the AR. (2) an implosion of peripheral field can happen simultaneously with an eruption, helping us track the magnetic energy transfer through a flaring region. MHD simulations, as in \citet{ama2014}, might profitably be used to explore the field evolution, and probe the validity of these statements. 

We have emphasised the overall magnetic evolution associated with the eruption and implosion, and have not explored other aspects, such as why the filament instability happens in the first place, or why the overlying arcade disappears in AIA wavebands after its implosion. Our main conclusion is that, in this event, we can successfully unify aspects of three main ways to understand coronal magnetic instabilities, namely the metastable eruption model, the implosion conjecture, and the standard ``CSHKP'' flare model, with the transfer of magnetic energy within the AR being central to the process.

\acknowledgments
The authors thank the anonymous referee and Natasha Jeffrey for helpful comments on the manuscript, and Yuhong Fan, Sarah Gibson, Laurel Rachmeler, Yihua Yan for discussion on implosion. L.\,F. acknowledges support from STFC Consolidated Grant ST/L000741/1. The research leading to these results has received funding from the European Community's Seventh Framework Programme (FP7/2007-2013) under grant agreement No. 606862 (F-CHROMA). J.\,K.\,T.~acknowledges financial support by Austrian Science Fund (FWF): P25383-N27. H.\,H. acknowledges support from NASA under contract No. NAS 5-98033 for RHESSI at UC Berkeley. I.\,G.\,H. acknowledges support of a Royal Society University Research Fellowship. The authors are grateful to NASA/SDO, AIA, HMI, RHESSI and GOES science teams for data access.

{\it Facilities:} \facility{AIA (SDO)}, \facility{HMI (SDO)}, \facility{RHESSI}, \facility{GOES}.

\end{document}